\documentclass[aps,prd,preprintnumbers,superscriptaddress,nofootinbib,floatfix,twocolumn,notitlepage,amsmath,amssymb]{revtex4-2}

\usepackage[normalem]{ulem}
\usepackage{graphicx}
\usepackage{dcolumn}
\usepackage{xspace}
\usepackage[utf8]{inputenc}
\usepackage{float}
\usepackage{subfigure}
\usepackage{url}
\usepackage{color}
\usepackage[dvipsnames,table]{xcolor}
\usepackage{comment}
\usepackage{adjustbox}
\usepackage{placeins}
\usepackage{booktabs} 
\usepackage{colortbl} 

\def\beq#1\eeq{\begin{align}#1\end{align}}

\newcommand{\rds}{\ensuremath{{\cal R}(D^*)}}
\newcommand{\flell}{F_L^{\ell}}
\newcommand{\afbell}{A_{\rm FB}^{\ell}}
\newcommand{\vcb}{V_{cb}}

\definecolor{BlueViolet}{rgb}{0.2, 0.00, 0.7}
\definecolor{Blue}{rgb}{0.15, 0.00, 0.9}
\definecolor{halayaube}{rgb}{0.4, 0.22, 0.33}
\definecolor{sanddune}{rgb}{0.59, 0.44, 0.09}

\newcommand{\be}{\begin{equation}}
\newcommand{\ee}{\end{equation}}
\newcommand{\bea}{\begin{eqnarray}}
\newcommand{\eea}{\end{eqnarray}}

\usepackage[colorlinks=true,linkcolor=Blue,citecolor=Blue,urlcolor=BlueViolet]{hyperref}
\usepackage[hyphenbreaks]{breakurl} 

\begin{document} 

\preprint{PSI-PR-23-15}
\preprint{ZU-TH 22/23}
\preprint{TTP23-019}
\preprint{P3H-23-033}
\preprint{LAPTH-020/23}

\title{\boldmath 
Discriminating \texorpdfstring{$B\to D^{*}\ell\nu$}{B to D* ell nu} form factors via polarization observables and asymmetries
}

\author{Marco Fedele}
\email[]{marco.fedele@kit.edu}
\affiliation{Institut f\"ur Theoretische Teilchenphysik (TTP), Karlsruhe Institute of Technology, D-76131 Karlsruhe, Germany}

\author{Monika Blanke} 
\email[]{monika.blanke@kit.edu}
\affiliation{Institut f\"ur Theoretische Teilchenphysik (TTP), Karlsruhe Institute of Technology, D-76131 Karlsruhe, Germany}
\affiliation{Institut f\"ur Astroteilchenphysik (IAP),
Karlsruhe Institute of Technology, D-76344 Eggenstein-Leopoldshafen, Germany}

\author{Andreas Crivellin} 
\email[]{andreas.crivellin@cern.ch}
\affiliation{Paul Scherrer Institut, CH–5232 Villigen PSI, Switzerland}
\affiliation{Physik-Institut, Universit\"at Z\"urich, Winterthurerstrasse 190, 8057 Z\"urich, Switzerland}

\author{\mbox{Syuhei Iguro}}
\email[]{igurosyuhei@gmail.com}
\affiliation{Institut f\"ur Theoretische Teilchenphysik (TTP), Karlsruhe Institute of Technology, D-76131 Karlsruhe, Germany}
\affiliation{Institut f\"ur Astroteilchenphysik (IAP),
Karlsruhe Institute of Technology, D-76344 Eggenstein-Leopoldshafen, Germany}

\author{Ulrich Nierste}
\email[]{ulrich.nierste@kit.edu}
\affiliation{Institut f\"ur Theoretische Teilchenphysik (TTP), Karlsruhe Institute of Technology, D-76131 Karlsruhe, Germany}

\author{Silvano Simula}
\email[]{silvano.simula@roma3.infn.it}
\affiliation{Istituto Nazionale di Fisica Nucleare, Sezione di Roma Tre, Via della Vasca Navale 84, I-00146 Rome, Italy}

\author{Ludovico Vittorio}
\email[]{ludovico.vittorio@lapth.cnrs.fr}
\affiliation{LAPTh, Université Savoie Mont-Blanc and CNRS, F-74941 Annecy, France}


\begin{abstract}
\noindent
Form factors are crucial theory input in order to extract $|V_{cb}|$ from $B \to D^{(*)}\ell\nu$ decays, to calculate the Standard Model prediction for ${\cal R}(D^{(*)})$ and to assess the impact of New Physics. In this context, the Dispersive Matrix approach, a first-principle calculation of the form factors, using no experimental data but rather only lattice QCD results as input, was recently applied to $B \to D^{(*)}\ell\nu$. It predicts (within the Standard Model) a much milder tension with the ${\cal R}(D^*)$ measurements than the other form factor approaches, while at the same time giving a value of $|V_{cb}|$ compatible with the inclusive value. However, this comes at the expense of creating tensions with differential $B\to D^*\ell\nu$ distributions (with light leptons). In this article, we explore the implications of using the Dispersive Matrix method form factors, in light of the recent Belle~(II) measurements of the longitudinal polarization fraction of the $D^*$ in $B\to D^*\ell\nu$ with light leptons, $F_L^{\ell}$, and the forward-backward asymmetry, $A_{\rm FB}^{\ell}$. We find that the Dispersive Matrix approach predicts a Standard Model value of $F_L^{\ell}$ that is in significant tension with these measurements, while mild deviations in $A_{\rm FB}^{\ell}$ appear. Furthermore, $F_L^{\ell}$ is very insensitive to New Physics such that the latter cannot account for the tension between Dispersive Matrix predictions and its measurement. While this tension can be resolved by deforming the original Dispersive Matrix form factor shapes within a global fit, a tension in ${\cal R}(D^*)$ reemerges. As this tension is milder than for the other form factors, it can be explained by New Physics not only in the tau lepton channel but also in the light lepton modes. 
\end{abstract}
\maketitle


\section{Introduction}

Due to the chiral suppression of purely leptonic $B$ decays, the most precise direct determinations of the CKM element $|\vcb|$ originate from semi-leptonic $B$ decays~\cite{Bigi:2016mdz,Bernlochner:2017jka,Jaiswal:2017rve,Bernlochner:2019ldg,Gambino:2019sif, Bordone:2019vic,Jaiswal:2020wer,Martinelli:2021onb,Bernlochner:2022ywh,Biswas:2022yvh,Cui:2023bzr,HFLAV:2022pwe}. However, the exclusive and inclusive determinations of $|\vcb|$ have been in tension with each other for a long time now~\cite{FlavourLatticeAveragingGroupFLAG:2021npn,HFLAV:2022pwe}. Furthermore, the exclusive value is also in tension with the indirect determination of $|\vcb|$ from the Unitarity Triangle analysis~\cite{UTfit:2022hsi,CKMfitter}, which points towards its inclusive value. As this inclusive/exclusive discrepancy cannot be explained by physics beyond the Standard Model (SM)~\cite{Crivellin:2014zpa,Jung:2018lfu,Iguro:2020cpg},\footnote{Previously, an explanation via a right-handed charged current was possible~\cite{Crivellin:2009sd}. However, with the recent exclusive values of $|\vcb|$ from $B\to D^{(*)} \ell \bar\nu$, which are both below the inclusive one, this is not feasible anymore.} scrutiny of the different theoretical methods is even more important. 

In the exclusive case, the SM value of $|\vcb|$ relies on form factors (FFs), which can be obtained by different non-perturbative methods, in particular light-cone sum rules~\cite{Ball:1992aa,Gubernari:2018wyi,Cui:2023bzr} and lattice QCD~\cite{FermilabLattice:2021cdg,Harrison:2023dzh}. These FFs also enter the SM predictions for the ratios
\begin{equation}\label{eq:RD_RDst}
{\cal R}(D^{(*)}) = \frac{{\rm BR}(B\to D^{(*)} \tau \bar\nu)}{{\rm BR}(B\to D^{(*)} \ell \bar\nu)}\,, \quad\quad \ell=e,\mu\,,
\end{equation}
testing lepton flavour universality (LFU). Here, a tension between the measurements of BaBar~\cite{BaBar:2012obs,BaBar:2013mob}, Belle~\cite{Belle:2015qfa,Belle:2016ure,Hirose:2016wfn,Hirose:2017dxl, Belle:2019rba} and LHCb~\cite{Aaij:2015yra,Aaij:2017uff,Aaij:2017deq,LHCbSem}, whose average reads~\cite{HFLAV:2022pwe}
\begin{eqnarray}
\begin{aligned}
{\cal R}(D)\,=\,0.356\pm0.029\,,\\
{\cal R}(D^*)\,=\,0.284\pm0.013\,,
\label{eq:HFLAV}
\end{aligned}
\end{eqnarray}
and the SM prediction~\cite{HFLAV:2022pwe}\footnote{The HFLAV value is based on Refs.~\cite{Bigi:2016mdz,Bernlochner:2017jka,Jaiswal:2017rve,Gambino:2019sif,Bordone:2019vic,Martinelli:2021onb}.}
 \begin{eqnarray}
\begin{aligned}
{\cal R}_{\rm SM}(D)\,=\,0.298\pm0.004 \,, \\
{\cal R}_{\rm SM}(D^*) \,=\,0.254\pm0.005 \,,
\label{eq:HFLAVSM}
\end{aligned} 
\end{eqnarray}
of $3.2\,\sigma$ exists. While in this case a New Physics (NP) explanation is possible, and these ratios have the advantage of being fairly insensitive to hadronic uncertainties, the HFLAV value is challenged by alternative calculations of the FFs, in particular the one based on the Dispersive Matrix (DM) approach~\cite{DiCarlo:2021dzg,Martinelli:2021frl}. The DM FFs do not only reduce the tension in $\mathcal{R}(D^{*})$ to $1.3\,\sigma$~\cite{Martinelli:2021myh}, but also give $|\vcb|=(41.2\pm 0.8)\times 10^{-3}$ from $B_{(s)}\to D_{(s)}^{(*)}\ell\nu$~\cite{Martinelli:2021onb,Martinelli:2021myh,Martinelli:2022xir}, which agrees with the inclusive values of $(42.16\pm0.51)\times10^{-3}$~\cite{Bordone:2021oof} or $(41.69\pm0.63)\times10^{-3}$~\cite{Bernlochner:2022ucr}.

However, this agreement comes at the cost of creating tensions between the DM FFs and the ones measured in differential $B\to D^{(*)}\ell\nu$ distributions. While this opens up the possibility of NP coupling to light leptons (instead or in addition to taus) it is not clear that this is a feasible option once all experimental information is taken into account. Indeed, the Belle and Belle~II collaborations recently released the results of the first measurement of the $D^*$ longitudinal polarization fraction $F_L^{\ell}$~\cite{Belle:2023bwv,BelleIISem} finding
\begin{eqnarray}
\begin{aligned}
F_{L,\, {\rm Belle}}^{e}\,=\,0.485\pm0.017\pm0.005  \,,\\
F_{L,\, {\rm Belle}}^{\mu}\,=\,0.518\pm0.017\pm0.005 \,,\label{eq:FLBelle}\\
F_{L,\, {\rm Belle\, II}}^{e}\,=\,0.521\pm0.005\pm0.007  \,,\\
F_{L,\, {\rm Belle\, II}}^{\mu}\,=\,0.534\pm0.005\pm0.006 \,.\\
\end{aligned}
\end{eqnarray}
Here, the first uncertainties are statistical while the second ones are systematic. Moreover, also the forward-backward asymmetry $A_{\rm FB}^\ell$ was measured, for which they find
\begin{eqnarray}
\begin{aligned}
A_{\rm FB,\,Belle}^{e}\,=\,0.230\pm0.018\pm0.005  \,,\\
A_{\rm FB,\,Belle}^{\mu}\,=\,0.252\pm0.019\pm0.005 \,,\\
\label{eq:AFBBelle}
A_{\rm FB,\,Belle\, II}^{e}\,=\,0.219\pm0.011\pm0.020  \,,\\
A_{\rm FB,\,Belle\, II}^{\mu}\,=\,0.215\pm0.011\pm0.022 \,.
\end{aligned}
\end{eqnarray}
For the case of $F_L$, while a small tension is present among the two measurements in the electron channel, the muon ones are in good agreement. The situation is opposite for $A_{\rm FB}$, with the two measurements in the electron channel in good agreement, while a small discrepancy is present in the muon one. Importantly, the theoretical predictions for these quantities crucially depend on the choice of FF parameterization. 

Given this new level of accuracy in these observables, it is imperative to inspect the impact of these measurements on the global $b \to c \ell \nu$ fit, including the dependence on the FF set used. In this article, we  focus in particular on the DM FFs, compared to the FFs of Ref.~\cite{Bordone:2019vic,Gambino:2019sif,Iguro:2020cpg,FermilabLattice:2021cdg}. For this, we first review the formalism used to describe $B\to D^*\ell\nu$ decays in Sec.~\ref{sec:formalism} and give a short summary to the DM approach in Sec.~\ref{sec:dm}. Implications of the measurements of $F_L^{e,\mu}$ are discussed in Sec.~\ref{sec:implications} before we conclude in Sec.~\ref{sec:conclusions}.


\section{Formalism}\label{sec:formalism}

The effective Hamiltonian
\begin{equation}
\renewcommand{\arraystretch}{1.8}
\begin{array}{r}\label{eq:Heff}
 {\cal H}_{\rm eff}=  2\sqrt{2} G_{F} V^{}_{cb} \big[(1+g_{V_L}^{\ell}) O_{V_L}^{\ell} + g_{V_R}^{\ell} O_{V_R}^{\ell}
 \\   + g_{S_L}^{\ell} O_{S_L}^{\ell} + g_{S_R}^{\ell} O_{S_R}^{\ell}+g_{T}^{\ell} O_{T}^{\ell}\big] + \text{h.c.}\,,
\end{array}
\end{equation}
with the dimension-six operators
\begin{equation}
\renewcommand{\arraystretch}{1.8}
\begin{array}{l}
   O_{V_L}^{\ell}  = \left(\bar c\gamma ^{\mu } P_L b\right)  \left(\bar {\ell} \gamma_{\mu } P_L \nu_{{\ell}}\right)\,, \\ 
   O_{V_R}^{\ell}  = \left(\bar c\gamma ^{\mu } P_R b\right)  \left(\bar {\ell} \gamma_{\mu } P_L \nu_{{\ell}}\right)\,, \\ 
   O_{S_L}^{\ell}  = \left( \bar c P_L b \right) \left( \bar {\ell} P_L \nu_{{\ell}}\right)\,,   \\
   O_{S_R}^{\ell}  = \left( \bar c P_R b \right) \left( \bar {\ell} P_L \nu_{\ell}\right)\,, \\
   O_{T}^{\ell}  = \left( \bar c \sigma^{\mu\nu}P_L  b \right) \left( \bar {\ell} \sigma_{\mu\nu} P_L \nu_{{\ell}}\right)\,,   \\
\end{array}
\label{eq:Oeff}
\end{equation}
describes $\bar{B}\to D^*\ell\nu$ transitions within the SM and heavy NP extensions. Here $\sigma_{\mu\nu}=\frac i2 [\gamma_\mu,\gamma_\nu]$ and $P_{L,R} = (1 \mp \gamma_5)/2$, and we do not consider here the case of light right-handed neutrinos. Note that at the dimension-six level in the SMEFT, $g_{V_R}$ is lepton flavour-universal, implying $g_{V_R}^e=g_{V_R}^\mu=g_{V_R}^\tau$.

For the SM operator, the $B\to D^*$ matrix element is described as
\begin{eqnarray} 
\label{eq:matrix_el_Dstar}
&&\!\!\!\!\!\!\!\langle D^{*} (p,\epsilon)| \bar{c} \gamma^{\mu} P_L b |\bar{B}(p_{B})\rangle 
 = \\ 
 &&\!\!\!\!\!\!\!- \frac{V(q^{2})}{m_{B}+m_{D^{*}}}  \varepsilon^{\mu}_{\alpha \beta \gamma} \epsilon^{*\alpha}p^{\beta} q^{\gamma} + \, i\, A_0(q^2) \frac{m_{D^{*}}}{q^2} (\epsilon^{*} \!\cdot\! q ) q^{\mu}   \nonumber \\
&&\!\!\!\!\!\!\!\,  - \frac{i A_{1} (q^{2})}{2(m_{B}-m_{D^{*}})} \left[(m_{B}^2-m_{D^{*}}^2) \epsilon^{*\mu}-(\epsilon^{*} \!\cdot\! q) (p+p_{B})^{\mu}\right]  \nonumber \\
&&\!\!\!\!\!\!\!\,  - \, i A_{3}(q^{2}) \, \frac{ m_{D^{*}}}{q^{2}} (\epsilon^{*} \!\cdot\! q) \left[\frac{q^{2}}{m_{B}^{2}-m_{D^{*}}^{2}} (p+p_{B})^{\mu}-q^{\mu}\right]  \ ,\nonumber
\end{eqnarray}
with
\begin{equation}
\label{eq:A1A2A3}
2\,m_{D^{*}}A_{3}(q^{2}) = (m_{B}+m_{D^{*}}) A_{1}(q^{2})-(m_{B}-m_{D^{*}})A_{2}(q^{2}),
\end{equation}
where $q=p_B-p$, such that $q^2$ is the invariant mass of the dilepton pair. The FFs can be decomposed as
\begin{align}
\renewcommand{\arraystretch}{1.5}
\label{eq:ff_BtoDstar}
V(q^2) &= \dfrac{m_{B}+m_{D^{*}}}{2} g(w)  \, , \nonumber\\
A_{1}(q^2) &= \dfrac{f(w)}{m_{B} + m_{D^{*}}}  \, ,\\ 
A_{2}(q^2) &= \dfrac{1}{2} \dfrac{m_{B}+m_{D^{*}}}{(w^{2}-1)m_{B} m_{D^{*}}}\left[  
\left(w-\dfrac{m_{D^{*}}}{m_{B}}\right) f(w) - \dfrac{\mathcal{F}_{1}(w)}{m_{B}} \right] \, ,\nonumber\\  
A_{0}(q^2) &= \dfrac{1}{2} \dfrac{m_{B}+m_{D^{*}}}{\sqrt{m_{B} m_{D^{*}}}} P_1(w) \, , \nonumber
\end{align}
in the Boyd-Grinstein-Lebed (BGL) formalism~\cite{Boyd:1995cf, Boyd:1995sq,Boyd:1997kz} with
\begin{equation}
w = \frac{m_{B}^2+m_{D^{*}}^2 - q^2}{2m_{B}m_{D^{*}}}\,.
\end{equation}
The FFs obey two kinematical constraints: at zero recoil ($w=1$), where only two out of the three helicity amplitudes are independent when the $D^*$ meson is at rest,
\begin{equation}
\label{eq:KC1}
\mathcal{F}_1(1) = (m_B-m_{D^*})f(1)
\end{equation}
holds, while at maximum recoil, due to the cancellation of any apparent kinematical singularity in the Lorentz decomposition of Eq.~\eqref{eq:matrix_el_Dstar}, we have
\begin{equation}
\label{eq:KC2}
P_1 (w_{\rm max}) = \frac{\mathcal{F}_1(w_{\rm max})}{(1+w_{\rm max})(m_B-m_{D^*}) \sqrt{m_B m_{D^*}}}\,.
\end{equation}
Here, we introduced $w_{\rm max}$ which for a massless lepton reads
\begin{equation}
w_{max} = \frac{m_B^2 + m_{D^*}^2}{2 m_B m_{D^*}} \simeq 1.504\,.
\end{equation}

We can now define the differential decay width as
\begin{widetext}
\begin{equation}
\begin{aligned}
\label{eq:dGamma}
&\frac{d\Gamma(B \rightarrow D^{*}(\rightarrow D\pi) \ell \nu)}{dw\,d\!\cos \theta_{\ell}\, d\!\cos \theta_D\, d\!\chi} = \frac{G_F^2 \vert V_{cb} \vert^ 2}{4(4\pi)^4} 3 m_B m_{D^*}^2 \sqrt{w^2-1} \left( 1 - 2 \frac{m_{D^*}}{m_B} w + \frac{m_{D^*}^2}{m_B^2} \right) B(D^{*} \rightarrow D\pi) \\
&\hskip 3.98truecm \Big\{ (1-\cos \theta_{\ell} )^2 \sin^2 \theta_D \vert H_{+} \vert^2 + (1+\cos \theta_{\ell} )^2\sin^2 \theta_D \vert H_{-} \vert^2+ 4 \sin^2 \theta_{\ell}\cos^2 \theta_D\vert H_{0} \vert^2 \\
&\hskip 3.98truecm - 2 \sin^2 \theta_{\ell}\sin^2 \theta_D \cos 2\chi  H_{+}  H_{-} - 4 \sin \theta_{\ell} (1-\cos \theta_{\ell} ) \sin\theta_D\cos\theta_D\cos\chi H_{+}  H_{0} \\
&\hskip 3.98truecm + 4 \sin \theta_{\ell} (1+\cos \theta_{\ell} ) \sin\theta_D\cos\theta_D\cos\chi H_{-}  H_{0} \Big\},
\end{aligned}
\end{equation}
\end{widetext}
in the massless lepton limit.\footnote{In our analysis we nevertheless keep the full lepton mass dependence which introduces the additional FF $P_1$.} In the differential width, the angle (in the virtual $W$ boson rest frame) between the lepton and the direction opposite to the $B$ meson momentum is defined as $\theta_{\ell}$; the angle in the $D^{*}$ rest frame between the $D$ meson and the direction opposite the $B$ meson momentum is $\theta_D$, and finally the angle in the $B$ meson rest frame between the two decay planes spanned by the $D^{*} - D$ and $W - \ell$ systems is $\chi$. 

In Eq.~\eqref{eq:dGamma} we have introduced the helicity amplitudes
\begin{eqnarray}
\begin{aligned}
\label{helampl}
H_0(w) = \frac{\mathcal{F}_1(w)}{\sqrt{m_B^2+m_{D^*}^2-2m_Bm_{D^*}w}}\,,\\
H_{\pm}(w) = f(w) \mp m_B m_{D^*} \sqrt{w^2-1}\,g(w)\,,
\end{aligned}
\end{eqnarray}
which enter in the decay rate, in the $D^*$ longitudinal polarization fraction and in the forward-backward asymmetry:
\begin{eqnarray}
\frac{d\Gamma}{dw} \propto |H_0(w)|^2 + |H_+(w)|^2 + |H_-(w)|^2 \,,\label{eq:dGdw}\\
F_L^\ell(w) = \frac{|H_0(w)|^2}{|H_0(w)|^2 + |H_+(w)|^2 + |H_-(w)|^2}\,,\label{eq:FLw}\\
A_{\rm FB}^\ell(w) = \frac{|H_-(w)|^2 - |H_+(w)|^2}{|H_0(w)|^2 + |H_+(w)|^2 + |H_-(w)|^2}\,.\label{eq:AFBw}
\end{eqnarray}
These quantities are usually predicted and measured after integrating the helicity amplitudes over the available phase space, i.e.~$w \in [1,w_{\rm max}]$. This implies an interesting feature for the $\mathcal{F}_1(w)$ FF: if its integral over the phase space increases, e.g.~due to an increase of $\mathcal{F}_1(w)$ at large $w$, this induces an increase for both observables in Eqs.~\eqref{eq:dGdw}-\eqref{eq:FLw} and a decrease for the one in Eq.~\eqref{eq:AFBw}. Remembering now the definition of $\rds$ in Eq.~\eqref{eq:RD_RDst}, we observe that an increase in the light lepton $d\Gamma/dw$ induces a decrease in the LFUV ratio. We can therefore infer that a change of the shape of $\mathcal{F}_1(w)$, resulting in an increase (decrease) of its integrated value, would imply an increase (decrease) for the prediction for $F_L^\ell$ and a decrease (increase) for the ones of $\rds$ and $A_{\rm FB}^\ell$. As we will see in Sec.~\ref{sec:implications}, this is crucial for understanding how the DM FFs affect the predictions for $F_L^{\ell}$ and $A_{\rm FB}^{\ell}$ within the SM. 


\section{Summary of Form Factors}\label{sec:dm}

We give a short description of the four FF sets we use.\footnote{We do not include the CLN parametrization~\cite{Caprini:1997mu}, obtained by applying heavy quark symmetry to FFs based on the Heavy Quark Effective Theory (HQET)~\cite{Isgur:1989vq,Neubert:1993mb} due do several issues: first, an inconsistent treatment of FF ratios when going beyond zero recoil in its experimental implementations (which was however not originally present in Ref.~\cite{Caprini:1997mu}) led to inconsistent results already at first order in HQET~\cite{Bernlochner:2017jka}; moreover, the constraint between the slope and curvature of the leading Isgur-Wise function are problematic; finally, this method uses approximations such that it is e.g.~not possible to consistently incorporate $\Lambda_{\rm {QCD}}^2/m_c^2$ correction which are non-negligible~\cite{Jung:2018lfu}.} 

\begin{itemize}
    \item DM: See detailed explanation below;
    \item F/M: The results obtained employing the lattice QCD by the Fermilab(FNAL)/MILC collaboration~\cite{FermilabLattice:2021cdg}, which for the first time computed the FFs at non-zero recoil and analysed them with the BGL parametrization, which is model-independent and based only on QCD dispersion relations\footnote{Novel results have recently been released by the HPQCD collaboration as well~\cite{Harrison:2023dzh}, compatible at the $1\sigma$ level with the FNAL/MILC results. We do, however, not consider this set in our analysis, because, on the one hand, it is not published yet, and, on the other hand, the DM results are based on the FNAL/MILC results only, which are therefore the natural lattice results to compare with.};
    \item BGJS: The recent results by Bigi, Gambino, Jung and Schacht~\cite{Bigi:2017njr,Bigi:2017jbd,Gambino:2019sif} without the inclusion of FNAL/MILC data~\cite{FermilabLattice:2021cdg} obtained employing the BGL parametrization~\cite{Boyd:1997kz} at the (2,2,2) truncation order; the additional pseudoscalar FF $P_1$ entering in $R(D^*)$ is extracted rescaling the FF $A_1$ by an appropriate ratio, computed in HQET~\cite{Bernlochner:2017jka}, and making use of the constraint~\eqref{eq:KC2}, which applies at maximum recoil;\footnote{A modified BGL expansion, which is characterized by a built-in unitarity condition, has recently been proposed in Ref.~\cite{Flynn:2023qmi}. However, we do not use it in this work since the results of its direct application to semileptonic $B \to D^*$ decays are not available at present.}
    \item IgWa: An approach developed in Refs.~\cite{Jung:2018lfu,Bordone:2019vic} to go beyond the original formulation of HQET~\cite{Isgur:1989vq,Neubert:1993mb}, that adopts a systematic expansion in terms of inverse powers of heavy quark masses and in particular incorporates $\Lambda_{\rm {QCD}}^2/m_c^2$ corrections coming from Heavy Quark symmetry considerations~\cite{Falk:1992wt}; in our numerical study we employ the latest results by Iguro and Watanabe based on their $(2/1/0)$ truncation order~\cite{Iguro:2020cpg}; we stress that these method employs as external input information coming from light-cone sum rules~\cite{Gubernari:2018wyi} and QCD sum rules~\cite{Neubert:1992wq,Neubert:1992pn,Ligeti:1993hw} as well, in order to constrain its parameters.
    \end{itemize}

Let us now consider the DM formalism in some more detail, as these FFs have never been used before in a global fit. The DM approach is a non-perturbative method for computing hadronic FFs in a model-independent way~\cite{DiCarlo:2021dzg}, applied, in particular, to the investigation of $B \to D^{(*)} \ell \nu_\ell$~\cite{Martinelli:2021onb,Martinelli:2021myh}, $B_s \to D_s^{(*)} \ell \nu_\ell$~\cite{Martinelli:2022xir} and $B(B_s) \to \pi(K) \ell \nu_\ell$~\cite{Martinelli:2022tte} decays.

The starting point is a dispersion relation that, for a generic FF $f$ can be written as~\cite{Boyd:1994tt,Boyd:1997kz,Caprini:1997mu} 
\be
\label{eq:JQ2z}
\frac{1}{2\pi i } \oint_{\vert z\vert =1} \frac{dz}{z}   \vert\phi(z) f(z)\vert^2 \leq \chi\, , 
\ee
where $\phi(z)$ is a function depending on the specific spin-parity channel (and including the Blaschke factors needed to remove sub-threshold bound-state poles) and $\chi$ is the so-called susceptibility related to the derivative of the Fourier transform of a suitable Green function of bilinear quark operators~\cite{Boyd:1997kz}, calculated within lattice QCD in Ref.~\cite{Martinelli:2021frl} for several spin-parity channels of interest. The conformal variable $z(t)$ is  defined as
\be
\label{eq:conformal}
z(t) = \frac{\sqrt{t_+ - t} - \sqrt{t_+ - t_-}}{\sqrt{t_+ - t} + \sqrt{t_+ - t_-}}\,,
\ee
with $t = q^2$ being the squared 4-momentum transfer and $t_\pm \equiv (m_B \pm m_{D^*})^2$ for the case of interest in this work.
By introducing the inner product~\cite{Bourrely:1980gp,Lellouch:1995yv}
\be
 \label{eq:inpro}
\langle g\vert h\rangle =\frac{1}{2\pi i } \oint_{\vert z\vert=1 } \frac{dz}{z}   \bar {g}(z) h(z)\, , 
\ee
where $\bar{g}(z)$ is the  complex conjugate of the function $g(z)$, Eq.~(\ref{eq:JQ2z}) can be also written as
\be
\label{eq:JQinpro}
0 \leq \langle \phi f \vert \phi  f\rangle \leq \chi\, .
\ee
Following Refs.~\cite{Bourrely:1980gp,Lellouch:1995yv} we introduce the set of functions
\be
g_t(z) \equiv \frac{1}{1-\bar{z}(t) z}\, ,
\ee
where $\bar{z}(t)$ is the complex conjugate of the conformal variable $z(t)$, so that the use of  Cauchy's theorem yields
\bea
\langle g_t|\phi f \rangle  & = & \phi(z(t))\, f\left(z(t)\right)\, ,  \\[2mm]
\langle g_{t_m} | g_{t_l} \rangle  & = & \frac{1}{1- \bar{z}(t_l) z(t_m)} \, .
\eea
The central ingredient of the DM method is the matrix~\cite{Bourrely:1980gp,Lellouch:1995yv}
\be
\label{eq:Delta}
{\tiny
\mathbf{M} \equiv \left(
\begin{array}{ccccc}
\langle\phi f | \phi f \rangle  & \langle\phi f | g_t \rangle  & \langle\phi f | g_{t_1} \rangle  &\cdots & \langle\phi f | g_{t_N}\rangle  \\[2mm]
\langle g_t | \phi f \rangle  & \langle g_t |  g_t \rangle  & \langle  g_t | g_{t_1} \rangle  &\cdots & \langle g_t | g_{t_N}\rangle  \\[2mm]
\langle g_{t_1} | \phi f \rangle  & \langle g_{t_1} | g_t \rangle  & \langle g_{t_1} | g_{t_1} \rangle  &\cdots & \langle g_{t_1} | g_{t_N}\rangle  \\[2mm]
\vdots & \vdots & \vdots & \vdots & \vdots \\[2mm] 
\langle g_{t_N} | \phi f \rangle  & \langle g_{t_N} | g_t \rangle  & \langle g_{t_N} | g_{t_1} \rangle  &\cdots & \langle g_{t_N} | g_{t_N} \rangle 
\end{array} \right)  ~ , ~
}
\ee
where $t_1, \ldots, t_N$ are the values of the squared 4-momentum transfer at which the FF $f(z)$ is known. In the DM method we consider only values $f(z(t_i))$ (with $i = 1, 2, ... N$) computed nonperturbatively on the lattice. 

The important feature of the matrix $\mathbf{M}$ is that, thanks to the positivity of the inner products, its determinant is positive semidefinite, i.e.\,$\det \mathbf{M} \geq 0$. This property is not modified when the first matrix element in Eq.~\eqref{eq:Delta} is replaced by the upper bound given by the susceptibility $\chi$ through Eq.~\eqref{eq:JQinpro}. Thus, using also the fact that both $z$ and $f(z)$ can assume only real values in the allowed kinematical region for semileptonic decays, the original matrix~\eqref{eq:Delta} can  be replaced  by
\be
\label{eq:Delta2}
{\small
\mathbf{M}_{\chi} = \left( 
\begin{tabular}{ccccc}
   $\chi$ & $\phi f$                            & $\phi_1 f_1$                             & $...$ & $\phi_N f_N$ \\[2mm]
   $\phi f$     & $\frac{1}{1 - z^2}$     & $\frac{1}{1 - z z_1}$      & $...$ & $\frac{1}{1 - z z_N}$ \\[2mm]
   $\phi_1 f_1$ & $\frac{1}{1 - z_1 z}$  & $\frac{1}{1 - z_1^2}$     & $...$ & $\frac{1}{1 - z_1 z_N}$ \\[2mm]
   $... $  & $...$                           & $...$                              & $...$ & $...$ \\[2mm]
   $\phi_N f_N$ & $\frac{1}{1 - z_N z}$ & $\frac{1}{1 - z_N z_1}$ & $...$ & $\frac{1}{1 - z_N^2}$
\end{tabular}
\right) ~ , ~
}
\ee
where $\phi_i f_i \equiv \phi(z_i) f(z_i)$ (with $i = 1, 2, ... N$) represent the known values of $\phi(z) f(z)$ corresponding to the given set of values $z_i$.
\begin{figure*}[!th!]
\centering
\includegraphics[width =0.195\textwidth]{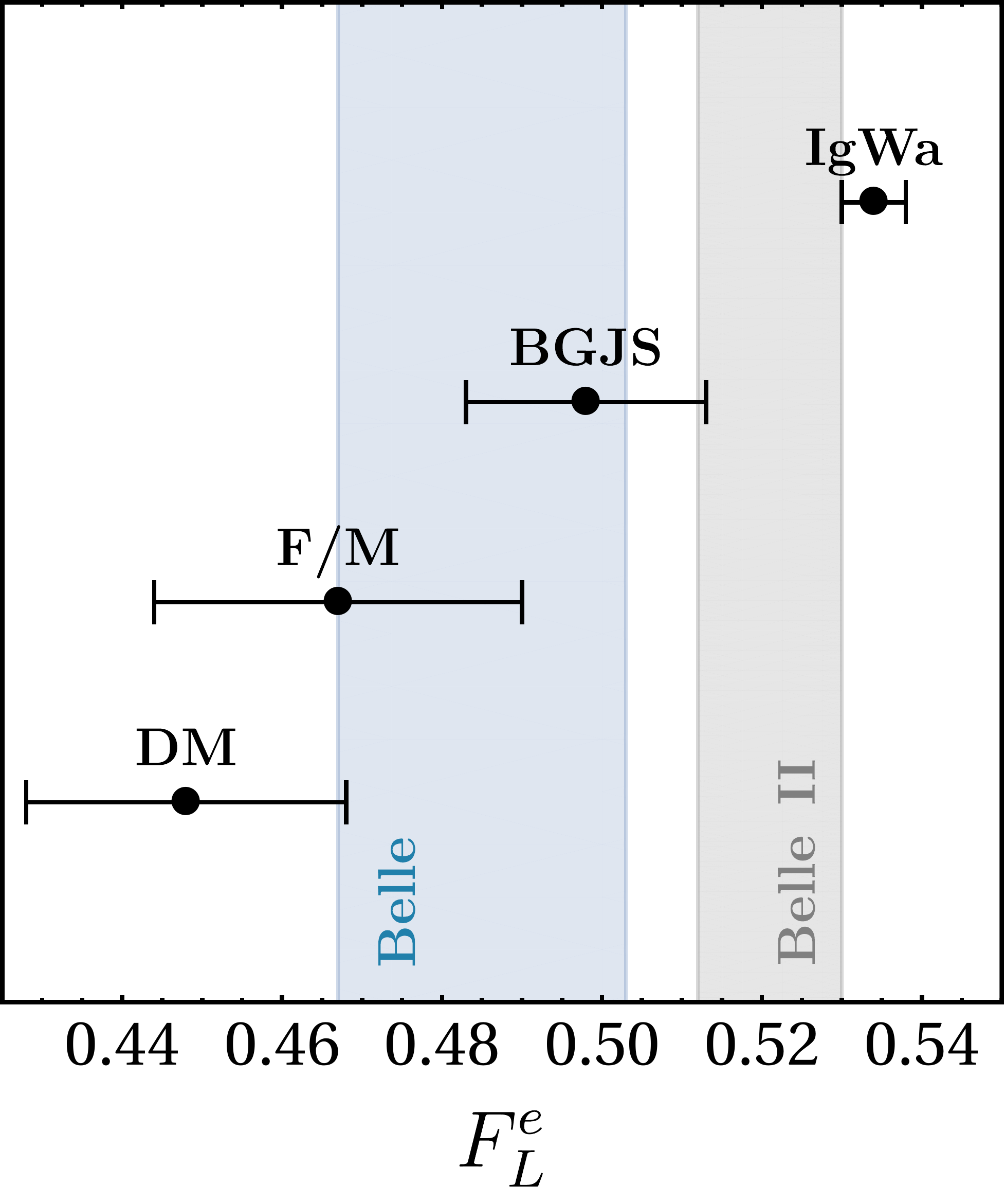}
\includegraphics[width =0.195\textwidth]{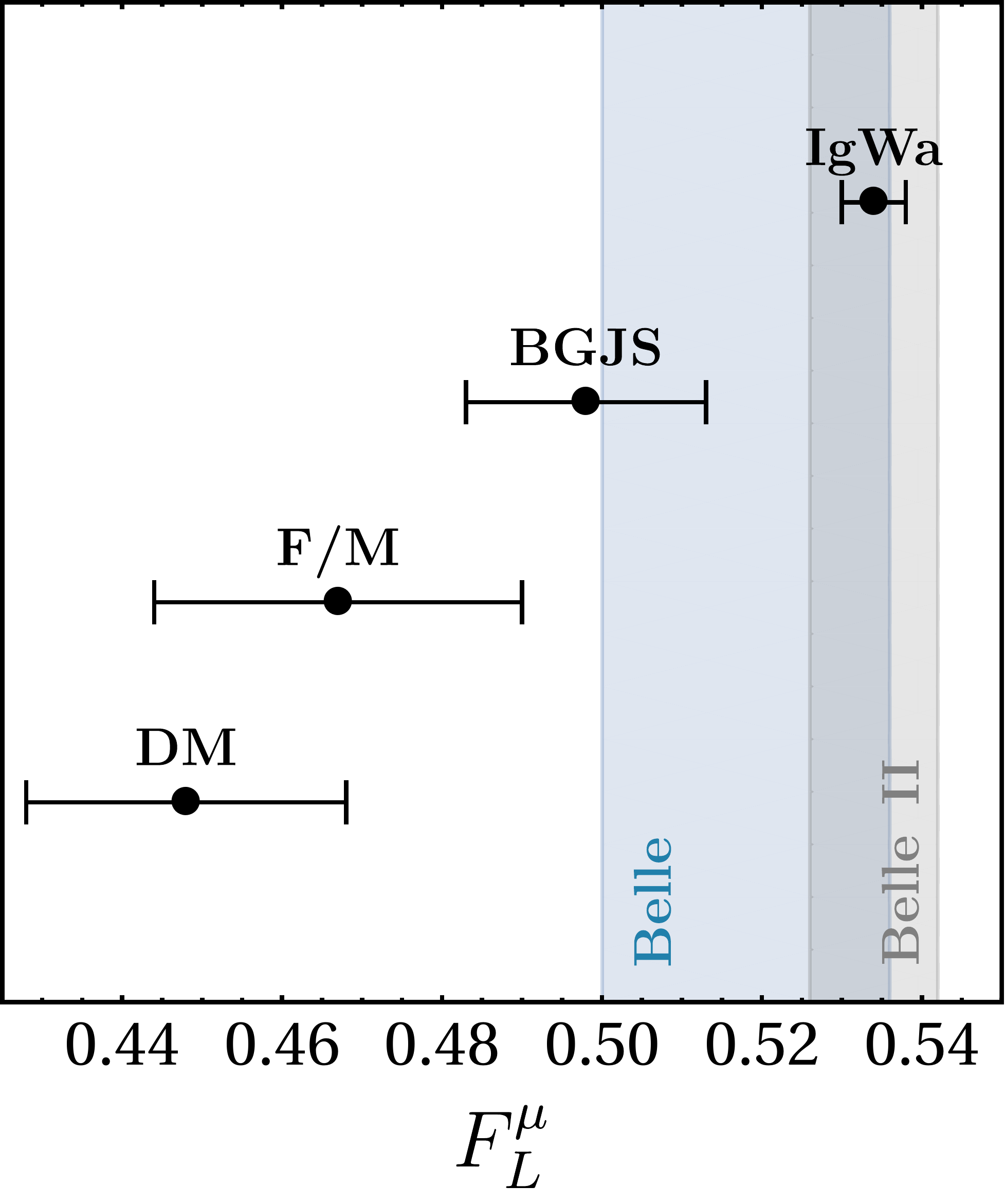}
\includegraphics[width =0.195\textwidth]{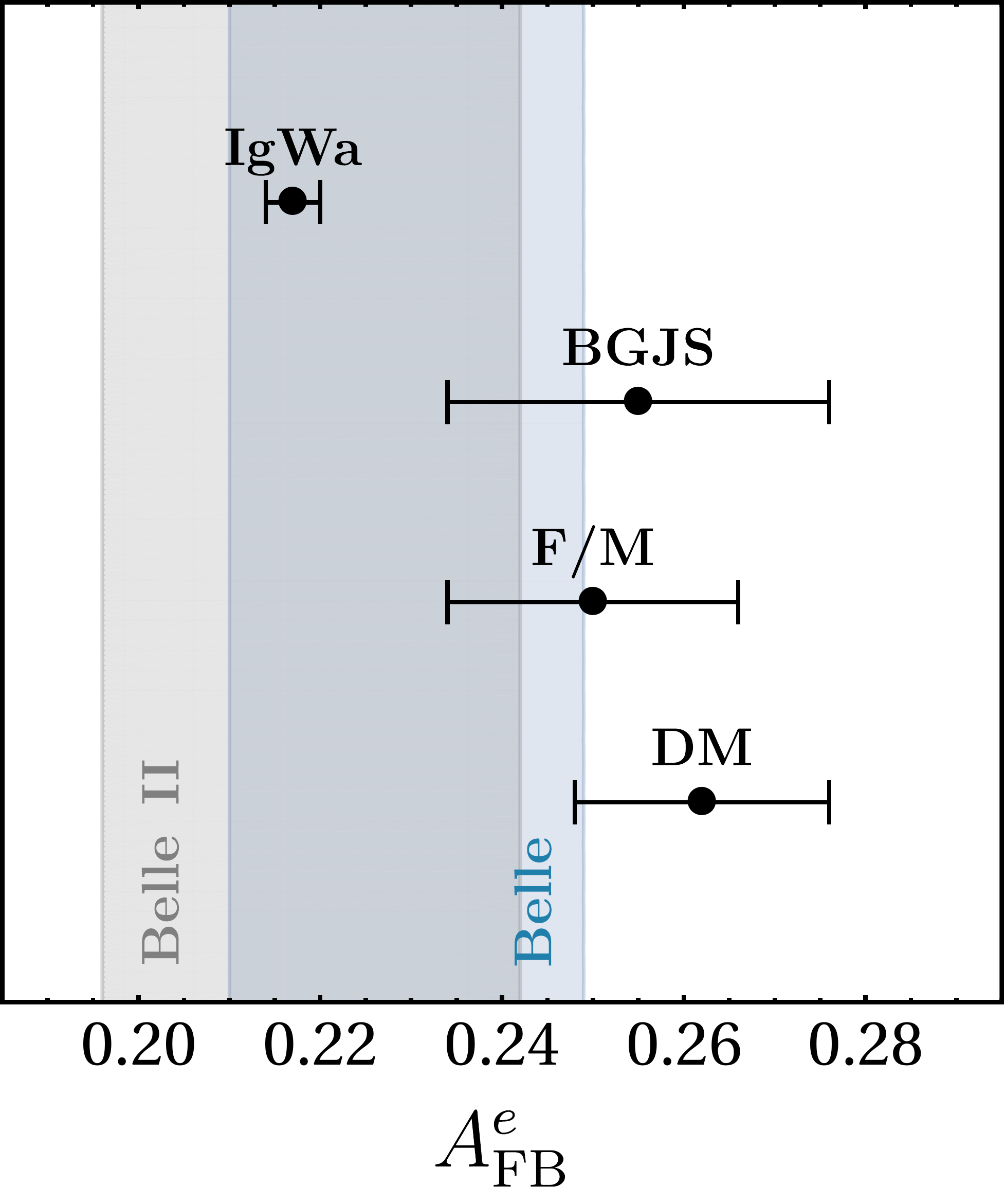}
\includegraphics[width =0.195\textwidth]{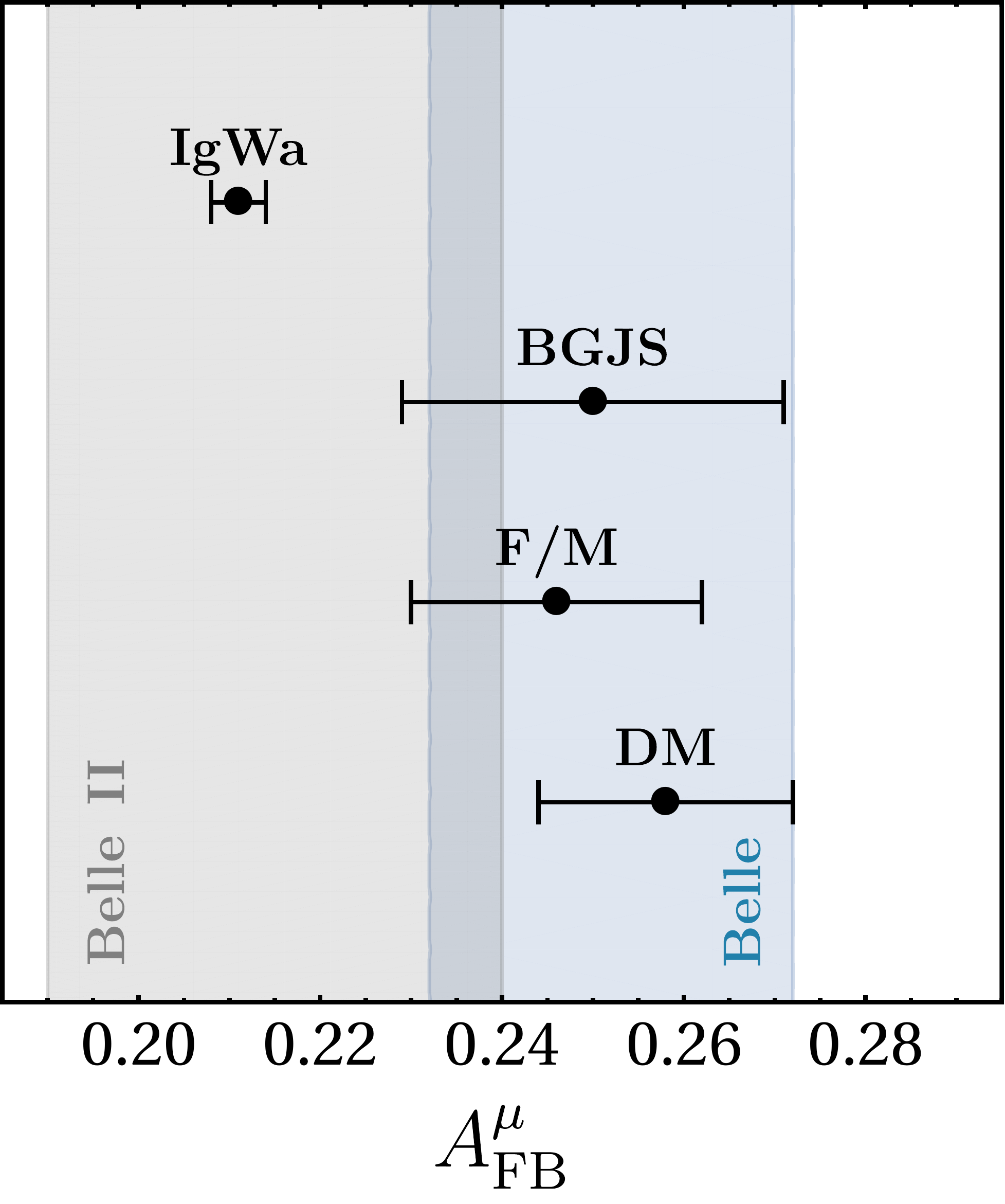}
\includegraphics[width =0.195\textwidth]{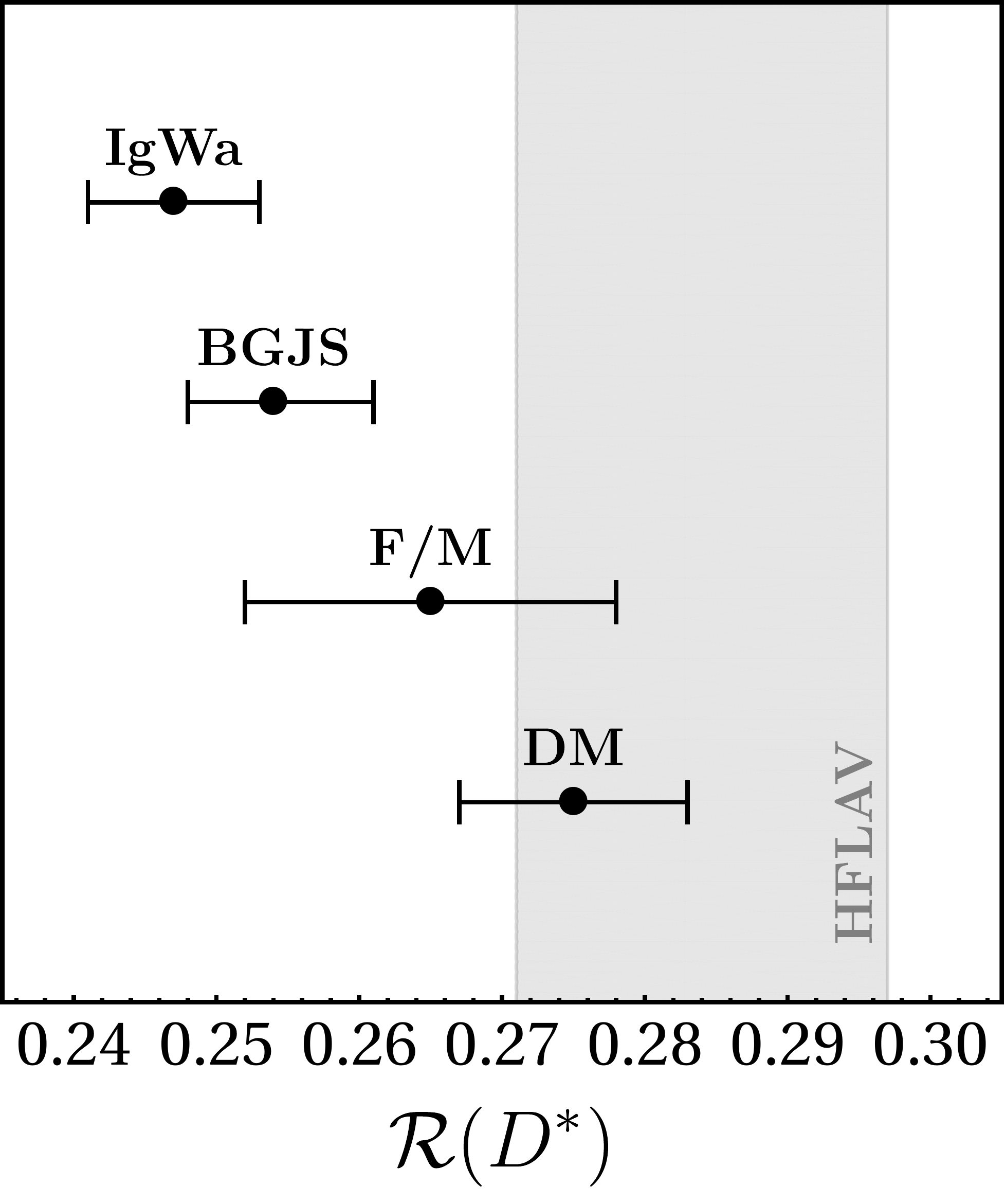}
\caption{Comparisons among measurements~\cite{Belle:2023bwv, BelleIISem, HFLAV:2022pwe} and SM predictions for $\flell$, $\afbell$ and $\rds$ at the $1\,\sigma$ level. See Sec.~\ref{subsec:sm} for details on the different predictions.
}
\label{fig:comparison}
\end{figure*}

By imposing the positivity of the determinant of the matrix~\eqref{eq:Delta2} it is possible to explicitly compute the lower and upper bounds that unitarity imposes on the FF $f(z)$ for a generic value of $z$, namely~\cite{DiCarlo:2021dzg}
\be
  \label{eq:bounds}
  \beta(z) - \sqrt{\gamma(z)} \leq f(z) \leq \beta(z) + \sqrt{\gamma(z)} ~ , ~
\ee 
where
\bea
      \label{eq:beta_final}
      \beta(z) & \equiv & \frac{1}{\phi(z) d(z)} \sum_{j = 1}^N \phi_j f_j d_j \frac{1 - z_j^2}{z - z_j} ~ , ~ \\
      \label{eq:gamma_final}
      \gamma(z) & \equiv &  \frac{1}{1 - z^2} \frac{1}{\phi^2(z) d^2(z)} \left( \chi - \chi_\text{DM} \right) ~ , ~ \\
      \label{eq:chi0_final}
      \chi_\text{DM} & \equiv & \sum_{i, j = 1}^N \phi_i f_i \phi_j  f_j d_i d_j \frac{(1 - z_i^2) (1 - z_j^2)}{1 - z_i z_j} ~ , ~ \\
      \label{eq:d0}
     d(z) & \equiv & \prod_{m = 1}^N \frac{1 - z z_m}{z - z_m}  ~ , ~ \\
     \label{eq:di}
     d_j & \equiv & \prod_{m \neq j = 1}^N \frac{1 - z_j z_m}{z_j - z_m}  ~ . ~ 
\eea
Unitarity is satisfied only when $\gamma(z) \geq 0$, which implies $\chi \geq \chi_\text{DM}$ and represents a parametrization-independent test of it for a given set of input values $f_j$. In this way, the input data are filtered by unitarity. Within the DM approach only the subset of input data satisfying the unitary filter $\chi \geq \chi_\text{DM}$ is considered.

Moreover, when $z \to z_j$ one has $\beta(z) \to f_j$ and $\gamma (z)\to 0$ (see Ref.~\cite{DiCarlo:2021dzg}). In other words, Eq.\,(\ref{eq:bounds}) exactly reproduces the input unitary data, i.e.\,the subset satisfying the unitary filter $\chi \geq \chi_\text{DM}$. Therefore, the DM band for the form factor $f(z)$ at a generic value of $z$ is given by the convolution of the uniform distribution corresponding to Eqs.~(\ref{eq:beta_final})-(\ref{eq:gamma_final}) with the distribution of the input (lattice) data $\{ f_j \}$ having $\chi \geq \chi_\text{DM}$. It represents the envelope of the results of all possible (either truncated or not truncated) $z$-expansions, like the BGL one~\cite{Boyd:1997kz}, which satisfy unitarity and at the same time exactly reproduce the input unitary data. In a frequentist language, this corresponds to a null value of the $\chi^2$-variable. This is at variance with what happens when working directly with (either truncated or not truncated) BGL fits, which may have $\chi^2 > 0$ even when the fitting function is constructed to satisfy unitarity.

For the purpose of our numerical analysis, for each of ten equally spaced recoil bins of the $B \to D^*$ channel we use a linear parametrization of the four DM FFs of the form
\begin{equation}
f_i(w) = a_{i} + w \cdot b_{i}\,.
\end{equation}
We stress that this is in not an expansion of the FFs in $w$, but rather a simple interpolation, bin by bin, of the full bands obtained in Ref.~\cite{Martinelli:2021myh}. The values of the parameters $a_i$ and $b_i$, their associated errors and their correlations (which we report in Appendix~\ref{sec:app}) are fixed by the reproduction of the values of the DM form factor $f_i(w)$ at the boundaries of the $i$-th experimental bin. Indeed, we have checked that with such a parametrization, the predicted values for $\rds$ and $F_L^\tau$~\cite{Martinelli:2021myh} are accurately reproduced. We then use values for each $a_{i}$ and $b_{i}$, assuming normally distributed errors and including correlations among the total of 80 values. The same procedure is adopted for the two FFs describing the $B \to D$ channel.

\section{Impact of \texorpdfstring{\boldmath $F_L^{\ell}$}{FLell} and \texorpdfstring{\boldmath $A_{\rm FB}^{\ell}$}{AFBell}}\label{sec:implications}

\subsection{SM predictions and fit}\label{subsec:sm}

In our analysis we distinguish between \emph{predictions} and \emph{fits}. In the former case, we only use the theoretical results relevant to each approach as priors in our analyses, without employing any further (experimental) information; in the latter case, all the relevant experimental results, like e.g.~Refs.~\cite{Belle:2023bwv,BelleIISem}, are enforced in the likelihood as well.

\begin{figure*}[!t!]
\centering
\includegraphics[width = 0.45\textwidth]{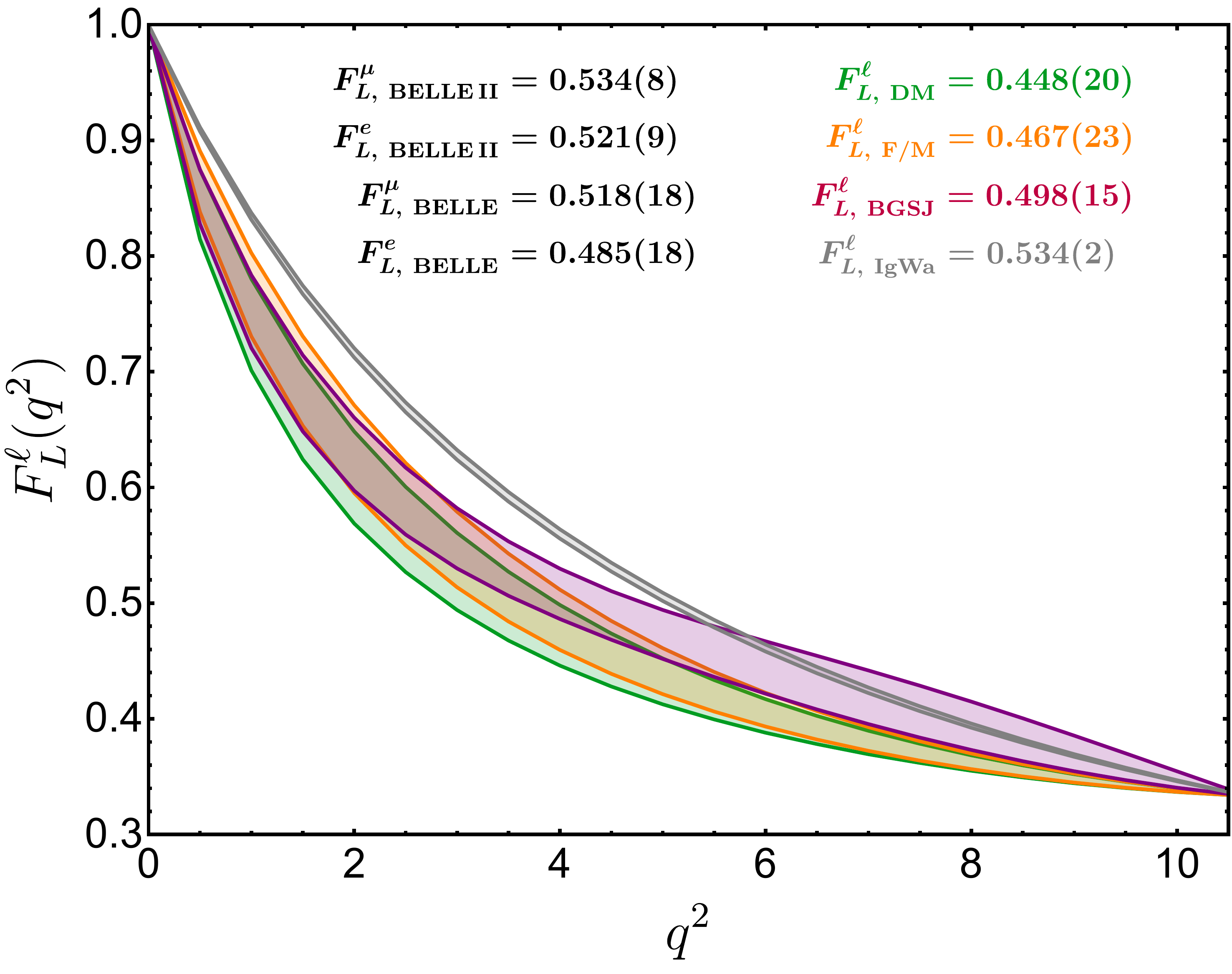}\hspace{2em}
\includegraphics[width = 0.45\textwidth]{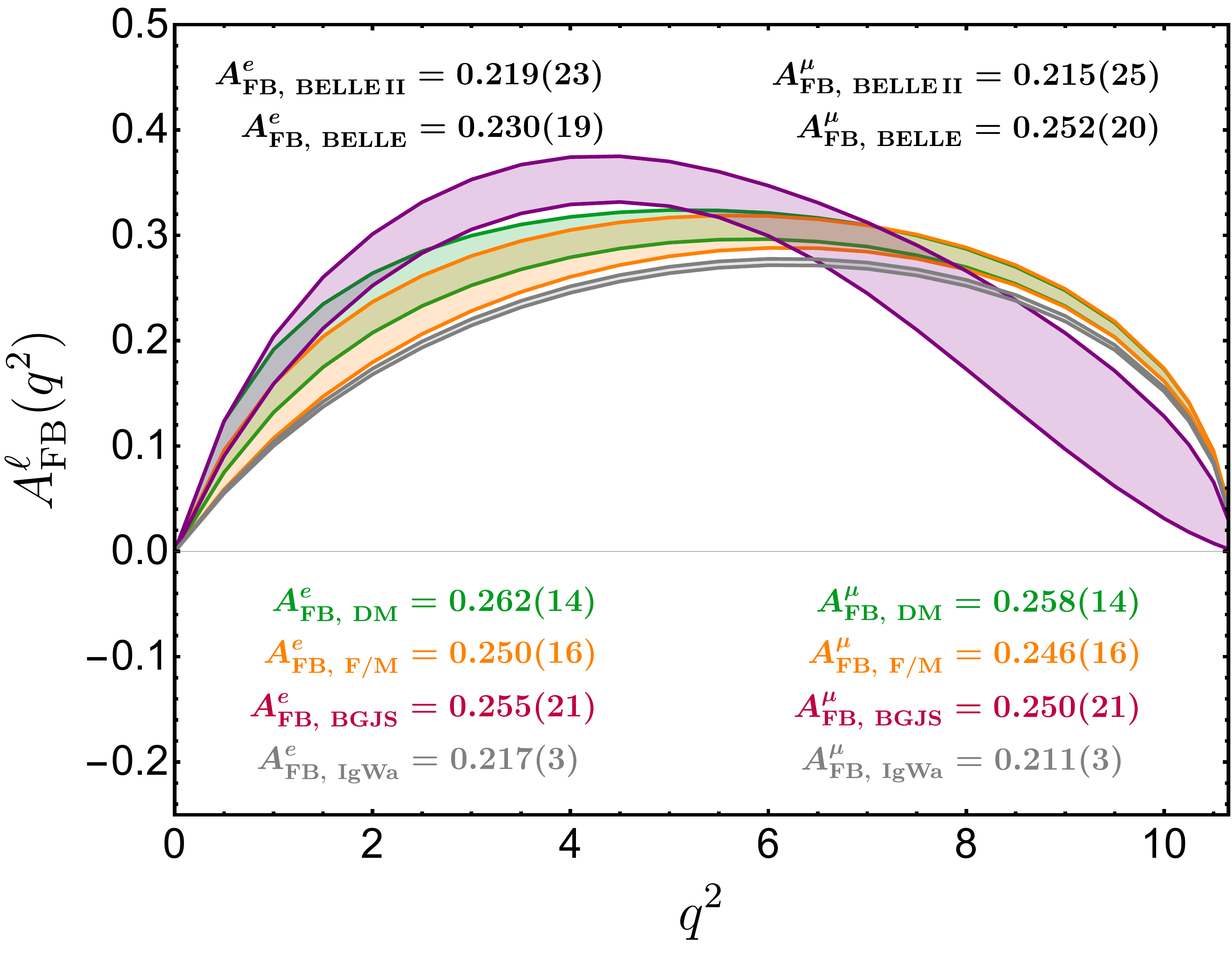}
\caption{Predicted $1\,\sigma$ range for $F_L^{\ell}$ (left panel) and $A_{\rm FB}^\ell$ (right panel) as a function of $q^2$ for the four different FF sets.}
\label{fig:FLAFB}
\end{figure*}

Let us start by studying $F_L^{\ell}$ and $A_{\rm FB}^{\ell}$ in the SM within the four sets of FFs discussed in the previous section. In the absence of NP, and without using any additional experimental information, $F_L^{\ell}$ ($\ell = e, \mu$) is predicted to be
\begin{eqnarray}
\begin{aligned}
F_{L,\, {\rm DM}}^{\ell}\,&=\,0.448\pm0.020  \,,\\
F_{L,\, {\rm F/M}}^{\ell}\,&=\,0.467\pm0.023 \,,\\
F_{L,\, {\rm BGJS}}^{\ell}\,&=\,0.498\pm0.015 \,,\\
F_{L,\, {\rm IgWa}}^{\ell}\,&=\,0.534\pm0.002 \,,
\label{eq:FLpreds}
\end{aligned}
\end{eqnarray}
which has to be compared with Eq.~\eqref{eq:FLBelle}. Similarly, the predictions for $A_{\rm FB}^{e,\mu}$ read
\begin{equation}
\begin{split}
A_{\rm FB,\, DM}^{e(\mu)}\,&=\,0.262(0.258)\pm0.014\,,\\
A_{\rm FB,\, F/M}^{e(\mu)}\,&=\,0.250(0.246)\pm0.016\\
A_{\rm FB,\, BGJS}^{e(\mu)}\,&=\,0.255(0.250)\pm0.021\,,\\
A_{\rm FB,\, IgWa}^{e(\mu)}\,&=\,0.217(0.211)\pm0.003\,,
\end{split}
\label{eq:AFBpreds}
\end{equation}
to be compared with Eq.~\eqref{eq:AFBBelle}. A visual summary can be found in Fig.~\ref{fig:comparison}. Note that at the current level of precision, there is no difference among the predictions of $F_L^e$ and $F_L^\mu$, while $A_{\rm FB}^e$ and $A_{\rm FB}^\mu$ slightly differ. 

For $\flell$ we observe in the DM approach compatibility in the electron channel with the Belle result but a $\sim2.5\,\sigma$ tension with Belle~II. In the muon channel, the tension is even $\sim2\,\sigma$ ($\sim3\,\sigma$) compared to the Belle (II) result. In the FNAL/MILC and BGJS approaches the tension is smaller, while the IgWa FFs describe data very well. Concerning $\afbell$, the DM approach predicts values slightly above both Belle~II measurements, while well reproducing the Belle ones. On the other hand, the IgWa FFs have a prediction lower than the Belle muon measurement but describe well the other data. Finally, both FNAL/MILC and BJSL results are in good agreement with all measured values.

To better understand these differences in the predictions of $\flell$ and $A_{\rm FB}^\ell$, it is useful to study their $q^2$ behaviour, as shown in the two panels of Fig.~\ref{fig:FLAFB}. The deviation between the DM and FNAL/MILC predictions originate from deviations in the shapes at smaller $q^2$. This behaviour is expected since the DM FFs use lattice input at low recoil, and a deviation among the two methods at large recoil can be traced back to the kinematic constraint in Eq.~\eqref{eq:KC2}, enforced only for the DM FFs. The values predicted in the BGJS approach deviate from the ones of the previous methods due to a difference in the shapes of the two observables. On the other hand, the IgWa results share the same shape with DM and FNAL/MILC while being shifted towards larger (lower) values for $\flell$ ($\afbell$), and display the smallest overall uncertainty.

Let us now focus on the DM FFs, which predict $\mathcal{R}(D^{*})$ to be in agreement with the measurements while having significant tensions in $\flell$ and, to a lesser extent, in $\afbell$. We perform a fit with HEPfit~\cite{DeBlas:2019ehy} in which we include the $\tau$ lepton polarization $P_\tau(D^*)$~\cite{Belle:2016dyj}, the longitudinal polarization fraction in $\tau$ decays $F_L^\tau$~\cite{Belle:2019ewo}, ${\cal R}(D^{(*)})$~\cite{HFLAV:2022pwe} and the longitudinal polarization fractions for light leptons $F_L^\ell$ and forward-backward asymmetries $A_{\rm FB}^\ell$~\cite{Belle:2023bwv,BelleIISem}.\footnote{We refrain from including  the differential results of Refs.~\cite{Belle:2017rcc,Belle:2018ezy} in our fit, given the potential incompatibility among the two results, see e.g. Ref.~\cite{Gambino:2019sif}.} The parameters we vary in the fit are the 80 $a_i$ and $b_i$ correlated FF parameters, for which we assume a multivariate gaussian prior, and whose central values, errors and correlations are reported in Appendix~\ref{sec:app}. We obtain the following results for all the relevant observables currently measured:
\begin{eqnarray}
\begin{aligned}
{\cal R}(D^*)_{\rm fit} \,&=\,0.265\pm0.005 \,,\\
F_{L,\, {\rm fit}}^{\ell}\,&=\,0.515\pm0.005 \,,\\
A_{\rm FB,\,fit}^{e}\,&=\,0.227\pm0.007 \,,\\
A_{\rm FB,\,fit}^{\mu}\,&=\,0.222\pm0.007 \,.
\label{eq:DMfit}
\end{aligned}
\end{eqnarray}
This means that the pull of the $F_L^\ell$ and $\afbell$ measurements on the shape of the FFs (for which, we remind, the SM values of the DM FFs act only as priors) are so strong that the post-fit values of the polarization fractions and asymmetries are compatible with data at the $1\,\sigma$ level, while a small tension at $1.5\,\sigma$ level in ${\cal R}(D^{*})$ reemerges. We, therefore, find ourselves in a situation similar to the other FF sets, i.e.~agreement within $F_L^\ell$ and $\afbell$ but tension in ${\cal R}(D^{*})$ (even though the latter is a bit less severe).

\begin{figure*}[!th!]
\centering
\includegraphics[width = 0.95\textwidth]{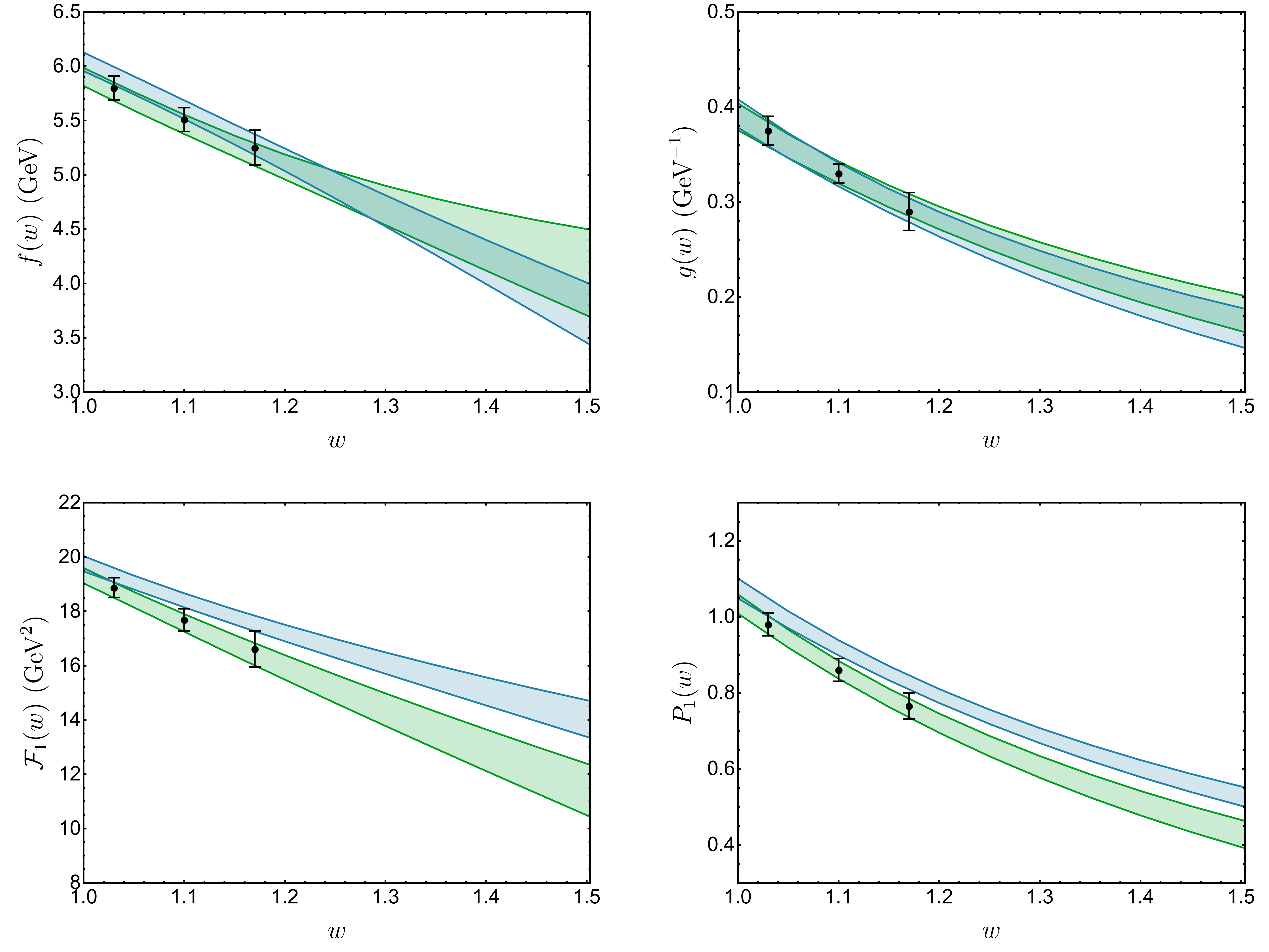}
\caption{$f(w)$ (top left panel), $g(w)$ (top right panel), $\mathcal{F}_1(w)$ (bottom left panel) and $P_1(w)$ (bottom right panel) in the DM approach when performing a prediction (green band) or a global fit (blue band); the black points and error bars denote the FNAL/MILC results~\cite{FermilabLattice:2021cdg}, which are used to calculate the green band. See Sec.~\ref{subsec:sm} for a discussion.}
\label{fig:FF}
\end{figure*}

In order to understand how the predicted values for $F_L^\ell$ and $\afbell$ in Eqs.~\eqref{eq:FLpreds}-\eqref{eq:AFBpreds} are related to the post-fit ones given in Eq.~\eqref{eq:DMfit}, it is useful to study the shape of the FFs in the two scenarios. We report them in Fig.~\ref{fig:FF}, together with the lattice data points~\cite{FermilabLattice:2021cdg} used in the DM approach. One sees that while the shape of $f(w)$ and $g(w)$ are not particularly altered when performing a global fit, the FFs $\mathcal{F}_1(w)$ and $P_1(w)$ show a $\sim30\%$ growth at $w \simeq 1.5$.\footnote{The changes in $\mathcal{F}_1(w)$ shape (and therefore in $P_1(w)$ one) is expected from Eq.~\eqref{eq:DMfit}, since an increase of its integrated value induces an enhancement of $\flell$ and a decrease of $\afbell$ and $\rds$, as detailed at the end of Sec.~\ref{sec:formalism}.} Moreover, the shape of $\mathcal{F}_1(w)$ and $P_1(w)$ are stretched to a point hardly compatible with the original DM values (pre-fit), not only at high recoil but also for lower values of $w$, where they now fail to reproduce lattice data. Hence, not only the tensions with data are present as in all other FF approaches, but in addition the post-fit FF shapes disagree with the lattice data used as input. Note that a similar tension among the experimental differential decay widths of Refs.~\cite{Belle:2017rcc,Belle:2018ezy} and the theoretical lattice data of Ref.~\cite{FermilabLattice:2021cdg} has already been pointed out in Refs.~\cite{FermilabLattice:2021cdg,Martinelli:2021myh}. 

Given this change in the FF shapes, it is interesting to study the implications regarding the extraction of $|\vcb|$. Going through the details of a systematic analysis of the correlated differential distributions from Refs.~\cite{Belle:2017rcc,Belle:2018ezy} is beyond the scope of this paper; however, as originally proposed in Ref.~\cite{Vittorio:2022rwi}, the CKM matrix element $|\vcb|$ can also be extracted from a comparison among the experimental determination of the total branching ratio and the corresponding theoretical value (modulo $|\vcb|^2$). The values of the total experimental branching ratio here considered are $\mathcal{B}(B \to D^* \ell \nu) = (4.95 \pm 0.11 \pm 0.22) \times 10^{-2}$~\cite{Belle:2017rcc} and $\mathcal{B}(B \to D^* \ell \nu) = (5.04 \pm 0.02 \pm 0.16) \times 10^{-2}$~\cite{Belle:2018ezy}. Following this procedure, the final value extracted for $|\vcb|$ corresponding to the FFs described by the green bands in Fig.~\ref{fig:FF} equals to $|\vcb|=(43.1\pm 1.2)\times 10^{-3}$, while the one induced by the FFs described by the blue bands in the same figure equals to $|\vcb|_{\rm fit}=(41.2\pm 1.2)\times 10^{-3}$, which is compatible with the inclusive determinations. It is also compatible with the value predicted by a Unitarity Triangle analysis (UTA), equal to $|\vcb|=(42.22\pm 0.51)\times 10^{-3}$~\cite{UTfit:2022hsi,CKMfitter} and obtained considering in the global analysis all relevant channels except the ones directly contributing to $|\vcb|$\footnote{The slopes of the FFs corresponding to the green bands in Fig.~\ref{fig:FF} do not agree with the experimental differential distributions by Belle~\cite{Belle:2017rcc,Belle:2018ezy}, as clearly visible from Fig.\,3 of Ref.~\cite{Martinelli:2021myh}. This {\it slope tension} is the reason of the FNAL/MILC result for $|\vcb|$, namely $|\vcb| = (38.4 \pm 0.8)\times 10^{-3}$, as given in Ref~\cite{FermilabLattice:2021cdg}. The difference with respect to our quoted value $|\vcb|=(43.1\pm 1.2)\times 10^{-3}$ is due to the fact that our procedure, instead, involves only integrals of the form factors over the total experimental decay range. In this sense, it is complementary to the one that can be obtained through a study of the experimental differential data.}. Let us finally highlight that systematic experimental uncertainties due to a proper estimation of the backgrounds and to the application of kinematical cuts (concerning in particular the lepton momentum) may arise at the percent level. These effects will have to be taken carefully into account especially in future analyses, namely when the precision on the theoretical computations of the FFs will importantly increase. 

\subsection{NP involving light leptons}

While it is well known that a deviation in $\rds$ can be explained by NP effects related to taus (see e.g.~Refs.~\cite{Fajfer:2012jt,Crivellin:2012ye} for very early accounts), it is interesting to see whether the predictions for $F_L^\ell$ and $\afbell$ with the DM FFs can be addressed by NP coupled to light leptons~\cite{Colangelo:2018cnj,Carvunis:2021dss}. In the following, due to the strong constraint from ${\cal R}_{e\mu}^{D*} = {\rm BR}(B\to D^{*} e \bar\nu)/{\rm BR}(B\to D^{*} \mu \bar\nu) = 0.990\pm0.031$~\cite{Belle:2023bwv} and ${\cal R}_{e\mu}^{D*}=1.001\pm0.023$~\cite{BelleIISem}, we assume LFU contributions to electrons and muons, namely $g_{i}^e=g_{i}^\mu\equiv g_{i}$.\footnote{We have checked that allowing for different NP effects in electrons and muons does not change the picture.}

Since left-handed vector operators only change the total decay width, we consider scalar and tensor operators as well as the right-handed vector current.\footnote{As explained in Sec.~\ref{sec:formalism}, we assume LFU contributions in all three leptons for the case of $g_{V_R}$.} 
However, when considering NP effects to one operator at a time, no significant preference for any non-vanishing Wilson coefficient is found.\footnote{Including NP effects in the scalar or tensor currents induces the presence of three additional matrix elements in the amplitude. In order to include the corresponding FFs, we rely on the results obtained in Ref.~\cite{Bernlochner:2017jka}.} The bounds at the $1\,\sigma$ level read
\begin{eqnarray}
\begin{aligned}
g_{V_R} \,&\in\,[-0.04,0.01] \,,\\
g_{S_L} \,&\in\,[-0.07,0.02] \,,\\
g_{S_R} \,&\in\,[-0.05,0.03] \,,\\
g_{T} \,&\in\,[-0.01,0.02] \,.
\label{eq:gNPres}
\end{aligned}
\end{eqnarray}
These results are not significantly changed if all coefficients are considered at the same time in a four-dimensional fit. This means that the tension between the $\flell$ measurements and the prediction using the DM FFs cannot be explained by NP as $\flell$ is very insensitive to it. This is in contrast to the case of NP coupled to tau leptons, where hints for scalar and/or tensor operators can be found, and $F_L^\tau$ and $A_{\rm FB}^\tau$ can be significantly altered by their presence (see e.g.~Ref.~\cite{Azatov:2018knx,Colangelo:2018cnj,Iguro:2022yzr} for recent reviews). This feature can be ascribed to the more accurate measurements for light lepton channels and the $m_\ell$ suppression in interference terms of scalar or tensor operators with the SM contribution.

If $\flell$ is included in the fit, a tension in ${\cal R}(D^{*})$ reemerges, but is still smaller than for the other FF sets. Therefore, it could be possible to explain these tensions with the SM operator $O_{V_L}^\ell$ related to light leptons (while for the other FF sets, a full explanation requires NP in taus). Performing a global fit in this setup to $g_{V_L}$ with light leptons, we obtain
\begin{eqnarray}
\begin{aligned}
{\cal R}(D)_{g_{V_L}} \,&=\,0.335\pm0.009 \,,\\
{\cal R}(D^*)_{g_{V_L}} \,&=\,0.291\pm0.009 \,,\\
F_{L,\, {g_{V_L}}}^{\ell}\,&=\,0.515\pm0.005 \,,\\
A_{{\rm FB},\,g_{V_L}}^{e}\,&=\,0.227\pm0.007 \,,\\
A_{{\rm FB},\,g_{V_L}}^{\mu}\,&=\,0.222\pm0.007 \,,
\label{eq:gVLfit}
\end{aligned}
\end{eqnarray}
corresponding to $g_{V_L}=-0.054\pm0.015$. As expected, $g_{V_L}$ induces only an overall normalization factor, and therefore both the longitudinal polarization fraction and the forward-backward asymmetry are insensitive to it, i.e.~their values stay unchanged compared to Eq.~\eqref{eq:DMfit}. Hence, the tensions in $\mathcal{F}_1(w)$ and $P_1(w)$ between their predictions within DM and the fitted values are still present. 

The presence of NP in $O_{V_L}^{\ell}$ alters the extraction of $|\vcb|$ by a factor of $1/(1+g_{V_L})$. According to our fit results, $|\vcb|_{\rm fit}=(41.2\pm 1.2)\times 10^{-3}$ from $B\to D^*\ell\nu$ would be shifted to $|\vcb|_{g_{V_L}}=(43.6\pm 1.4)\times 10^{-3}$. This value is still compatible with the inclusive one, which receives the same correction $1/(1+g_{V_L})$, and is not too large to spoil other indirect measurements like $\Delta F=2$ processes, sensitive to NP contributions. Finally, note that $g_{V_L}$ is small enough to evade constraints from high-$p_{\rm T}$ lepton tail searches~\cite{ATLAS:2019lsy} which require $|g_{V_L}|<0.25$~\cite{Iguro:2020keo}.

Importantly, an explanation via NP related to light leptons is only possible when employing the DM FFs in a global fit, albeit their significant deformation w.r.t.~their original shape. As stated above, such deformations points to a tension of lattice predictions with experiment, as also seen in the differential decay widths. On the other hand, the other FF approaches, due to the larger discrepancy in $\rds$, require larger values for $g_{V_L}$ which, if related to light leptons, would induce a shift in $|\vcb|$ too large to be compatible with the UTA and high-$p_{\rm T}$ constraints~\cite{Fedele:2022iib,Ray:2023xjn}.

\section{Conclusions}\label{sec:conclusions}

The SM predictions for $B\to D^{(*)}\ell\nu$ decays depend critically on the FFs used to calculate them. While most sets of FFs lead to significant tensions with the measurements of ${\cal R}(D^{*})$, the DM approach results in a SM value compatible with experiment. In order to further assess the agreement of the different FF sets with data, we investigated the impact of the recent measurements of the longitudinal $D^*$ polarization fraction $\flell$ for $\ell=e,\mu$ by Belle~\cite{Belle:2023bwv} and Belle II~\cite{BelleIISem}, as well as that of $A^\ell_{\rm FB}$. While FNAL/MILC, IgWa and BGJS lead to SM values in agreement with the experimental results, the DM predictions show significant tensions with them, up to the $\sim 3\,\sigma$ level. 

While it is well known that ${\cal R}(D^{*})$ can be accounted for by physics beyond the SM, we find that NP cannot explain the tension in $\flell$. The reason for this is that $\flell$ (with light leptons) is very insensitive to NP contributions, such that the constraints from other observables prevent a consistent NP explanation within a global fit. We then included $\flell$ within a global fit within the DM approach, using the theory predictions for the parameters as priors, and found that the pull on these parameters is so strong that it violently deforms them. One then ends up in a situation similar to the other three FF sets with a significant (unaccounted) tension in ${\cal R}(D^{(*)})$, which is however still smaller than in the other cases. 

This decreased tension allows for an explanation via $g_{V_L}$ with (only) light leptons (contrary to the other FFs where NP in tau leptons is needed), without causing problems in the determination of $|V_{cb}|$ or being in tension with direct LHC searches. However, a tension with the shape of the FFs predicted on the lattice is unavoidable, similar to the one observed previously~\cite{FermilabLattice:2021cdg,Martinelli:2021myh} when comparing lattice results~\cite{FermilabLattice:2021cdg} to experimental differential decay widths~\cite{Belle:2017rcc,Belle:2018ezy}.

\vspace{2mm} {\it Acknowledgments.}--- {\small The authors wish to thank Guido Martinelli and Manuel Naviglio for useful discussions, and Markus Prim and Chaoyi Lyu for fruitful exchanges about the recent Belle and Belle II results. The work of M.B., M.F., S.I.~and U.N.~is supported by the Deutsche Forschungsgemeinschaft (DFG, German Research Foundation) under grant 396021762-TRR 257. A.C.~is supported by a Professorship Grant (PP00P2\_176884) of the Swiss National Science Foundation. S.S.~is supported by the Italian Ministery of University and Research (MIUR) under grant PRIN20172LNEEZ. The work of L.V. is supported by ANR under contract n.~202650 (ANR-19-CE31-0016, GammaRare). U.N. acknowledges the hospitality of Fermilab and thanks Jim Simone and Andreas Kronfeld for helpful discussions.}

\appendix

\section{DM parameters and correlations}\label{sec:app}

In this Appendix, we provide to the reader the information necessary in order to reproduce the DM analyses performed in our study. As detailed in Sec.~\ref{sec:dm}, each FF is linearly parameterized by the form $f_i(w) = a_{i} + w \cdot b_{i}$, employing different values for the parameters $a_i$ and $b_i$ in ten recoil bins. In particular, the boundaries of the ten $w$ bins
are
\begin{equation}
    \{1.00, 1.05, 1.10, 1.15, 1.20, 1.25, 1.30, 1.35, 1.40, 1.45, 1.504\}\,.
\end{equation}

The numerical values and the associated errors of the $a_i$ parameters for the four FFs in the ten $w$ bins are given in Table~\ref{tab:ai_num}, while the values relative to the $b_i$ parameters are provided in Table~\ref{tab:bi_num}. Correlations among all the parameters are listed in Tables~\ref{tab:corr11}-\ref{tab:corr88}.

\begin{table}[h!]
\centering
\begin{tabular}{c|cccc}
\toprule
bin & $a_f$ & $a_g$ & $a_{\mathcal{F}_1}$ & $a_{P_1}$ \\
\hline
1 & $10.18(0.61)$ & $1.019(0.087)$ & $37.5(2.4)$ & $2.88(0.16)$ \\
2 & $9.95(0.59)$ & $0.953(0.078)$ & $37.1(2.3)$ & $2.68(0.14)$ \\
3 & $9.72(0.63)$ & $0.893(0.070)$ & $36.6(2.3)$ & $2.50(0.13)$ \\
4 & $9.49(0.70)$ & $0.837(0.064)$ & $36.2(2.5)$ & $2.34(0.12)$ \\
5 & $9.28(0.78)$ & $0.786(0.059)$ & $35.9(2.6)$ & $2.19(0.11)$ \\
6 & $9.08(0.87)$ & $0.739(0.054)$ & $35.5(2.8)$ & $2.05(0.10)$ \\
7 & $8.88(0.95)$ & $0.696(0.051)$ & $35.1(3.0)$ & $1.92(0.09)$ \\
8 & $8.69(1.03)$ & $0.656(0.048)$ & $34.8(3.2)$ & $1.80(0.08)$ \\
9 & $8.50(1.10)$ & $0.619(0.045)$ & $34.5(3.5)$ & $1.69(0.08)$ \\
10 & $8.32(1.16)$ & $0.584(0.043)$ & $34.2(3.8)$ & $1.59(0.07)$ \\
\bottomrule
\end{tabular}
\caption{\label{tab:ai_num} Numerical values and their associated errors for the $a_i$ parameters employed in our analyses.
}
\end{table}
\FloatBarrier

\begin{table}[h!]
\centering
\begin{tabular}{c|cccc}
\toprule
bin & $b_f$ & $b_g$ & $b_{\mathcal{F}_1}$ & $b_{P_1}$ \\
\hline
1 & $-4.26(0.58)$ & $-0.628(0.078)$ & $-18.2(2.3)$ & $-1.85(0.15)$ \\
2 & $-4.04(0.55)$ & $-0.565(0.069)$ & $-17.7(2.2)$ & $-1.66(0.13)$ \\
3 & $-3.83(0.57)$ & $-0.510(0.062)$ & $-17.4(2.3)$ & $-1.50(0.12)$ \\
4 & $-3.64(0.63)$ & $-0.462(0.056)$ & $-17.0(2.4)$ & $-1.36(0.11)$ \\
5 & $-3.46(0.69)$ & $-0.420(0.052)$ & $-16.7(2.5)$ & $-1.23(0.10)$ \\
6 & $-3.29(0.76)$ & $-0.382(0.048)$ & $-16.4(2.7)$ & $-1.12(0.09)$ \\
7 & $-3.14(0.81)$ & $-0.349(0.045)$ & $-16.1(2.8)$ & $-1.02(0.08)$ \\
8 & $-3.00(0.87)$ & $-0.319(0.043)$ & $-15.9(2.9)$ & $-0.93(0.08)$ \\
9 & $-2.87(0.91)$ & $-0.293(0.040)$ & $-15.6(3.1)$ & $-0.85(0.07)$ \\
10 & $-2.75(0.96)$ & $-0.269(0.039)$ & $-15.5(3.2)$ & $-0.78(0.07)$ \\
\bottomrule
\end{tabular}
\caption{\label{tab:bi_num} Numerical values and their associated errors for the $b_i$ parameters employed in our analyses.
}
\end{table}
\FloatBarrier

\begin{table}[h!]
\centering
{\footnotesize
\begin{tabular}{c|cccccccccc}
& $a_f^{1}$ & $a_f^{2}$ & $a_f^{3}$ & $a_f^{4}$ & $a_f^{5}$ & $a_f^{6}$ & $a_f^{7}$ & $a_f^{8}$ & $a_f^{9}$ & $a_f^{10}$ \\
\hline

$a_f^{1}$ & $1.$ & $0.94$ & $0.79$ & $0.64$ & $0.51$ & $0.41$ & $0.33$ & $0.27$ & $0.23$ & $0.19$ \\

$a_f^{2}$ & $0.94$ & $1.$ & $0.95$ & $0.87$ & $0.78$ & $0.7$ & $0.64$ & $0.59$ & $0.55$ & $0.51$ \\

$a_f^{3}$ & $0.79$ & $0.95$ & $1.$ & $0.98$ & $0.93$ & $0.88$ & $0.84$ & $0.8$ & $0.77$ & $0.74$ \\

$a_f^{4}$ & $0.64$ & $0.87$ & $0.98$ & $1.$ & $0.99$ & $0.96$ & $0.94$ & $0.91$ & $0.89$ & $0.87$ \\

$a_f^{5}$ & $0.51$ & $0.78$ & $0.93$ & $0.99$ & $1.$ & $0.99$ & $0.98$ & $0.97$ & $0.95$ & $0.94$ \\

$a_f^{6}$ & $0.41$ & $0.7$ & $0.88$ & $0.96$ & $0.99$ & $1.$ & $1.$ & $0.99$ & $0.98$ & $0.97$ \\

$a_f^{7}$ & $0.33$ & $0.64$ & $0.84$ & $0.94$ & $0.98$ & $1.$ & $1.$ & $1.$ & $0.99$ & $0.99$ \\

$a_f^{8}$ & $0.27$ & $0.59$ & $0.8$ & $0.91$ & $0.97$ & $0.99$ & $1.$ & $1.$ & $1.$ & $1.$ \\

$a_f^{9}$ & $0.23$ & $0.55$ & $0.77$ & $0.89$ & $0.95$ & $0.98$ & $0.99$ & $1.$ & $1.$ & $1.$ \\

$a_f^{10}$ & $0.19$ & $0.51$ & $0.74$ & $0.87$ & $0.94$ & $0.97$ & $0.99$ & $1.$ & $1.$ & $1.$ \\

\end{tabular}}
\caption{\label{tab:corr11} Correlations among the $a_f$ and $a_f$ parameters.
}
\end{table}
\FloatBarrier

\begin{table}[h!]
\centering
{\footnotesize
\begin{tabular}{c|cccccccccc}
& $a_g^{1}$ & $a_g^{2}$ & $a_g^{3}$ & $a_g^{4}$ & $a_g^{5}$ & $a_g^{6}$ & $a_g^{7}$ & $a_g^{8}$ & $a_g^{9}$ & $a_g^{10}$  \\
\hline

$a_f^{1}$ & $0.26$ & $0.26$ & $0.25$ & $0.23$ & $0.22$ & $0.2$ & $0.18$ & $0.15$ & $0.13$ & $0.1$ \\

$a_f^{2}$ & $0.33$ & $0.32$ & $0.31$ & $0.29$ & $0.27$ & $0.25$ & $0.23$ & $0.2$ & $0.17$ & $0.14$ \\

$a_f^{3}$ & $0.35$ & $0.34$ & $0.33$ & $0.32$ & $0.3$ & $0.28$ & $0.25$ & $0.22$ & $0.2$ & $0.16$ \\

$a_f^{4}$ & $0.34$ & $0.34$ & $0.33$ & $0.32$ & $0.3$ & $0.28$ & $0.25$ & $0.23$ & $0.2$ & $0.17$ \\

$a_f^{5}$ & $0.33$ & $0.33$ & $0.32$ & $0.31$ & $0.29$ & $0.27$ & $0.25$ & $0.22$ & $0.2$ & $0.17$ \\

$a_f^{6}$ & $0.32$ & $0.31$ & $0.3$ & $0.29$ & $0.28$ & $0.26$ & $0.24$ & $0.22$ & $0.19$ & $0.17$ \\

$a_f^{7}$ & $0.3$ & $0.3$ & $0.29$ & $0.28$ & $0.27$ & $0.25$ & $0.23$ & $0.21$ & $0.19$ & $0.16$ \\

$a_f^{8}$ & $0.29$ & $0.29$ & $0.28$ & $0.27$ & $0.26$ & $0.24$ & $0.23$ & $0.21$ & $0.18$ & $0.16$ \\

$a_f^{9}$ & $0.28$ & $0.28$ & $0.27$ & $0.26$ & $0.25$ & $0.24$ & $0.22$ & $0.2$ & $0.18$ & $0.16$ \\

$a_f^{10}$ & $0.27$ & $0.27$ & $0.27$ & $0.26$ & $0.25$ & $0.23$ & $0.21$ & $0.2$ & $0.18$ & $0.15$ \\

\end{tabular}}
\caption{\label{tab:corr12} Correlations among the $a_f$ and $a_g$ parameters.
}
\end{table}
\FloatBarrier

\begin{table}[h!]
\centering
{\footnotesize
\begin{tabular}{c|cccccccccc}
& $a_{\mathcal{F}_1}^{1}$ & $a_{\mathcal{F}_1}^{2}$ & $a_{\mathcal{F}_1}^{3}$ & $a_{\mathcal{F}_1}^{4}$ & $a_{\mathcal{F}_1}^{5}$ & $a_{\mathcal{F}_1}^{6}$ & $a_{\mathcal{F}_1}^{7}$ & $a_{\mathcal{F}_1}^{8}$ & $a_{\mathcal{F}_1}^{9}$ & $a_{\mathcal{F}_1}^{10}$  \\
\hline

$a_f^{1}$ & $0.72$ & $0.71$ & $0.65$ & $0.58$ & $0.52$ & $0.47$ & $0.42$ & $0.37$ & $0.33$ & $0.3$ \\

$a_f^{2}$ & $0.66$ & $0.67$ & $0.64$ & $0.59$ & $0.53$ & $0.49$ & $0.44$ & $0.4$ & $0.35$ & $0.31$ \\

$a_f^{3}$ & $0.54$ & $0.57$ & $0.56$ & $0.53$ & $0.49$ & $0.46$ & $0.42$ & $0.38$ & $0.33$ & $0.29$ \\

$a_f^{4}$ & $0.41$ & $0.47$ & $0.48$ & $0.46$ & $0.44$ & $0.41$ & $0.37$ & $0.34$ & $0.3$ & $0.26$ \\

$a_f^{5}$ & $0.32$ & $0.38$ & $0.4$ & $0.4$ & $0.38$ & $0.36$ & $0.33$ & $0.3$ & $0.27$ & $0.23$ \\

$a_f^{6}$ & $0.24$ & $0.31$ & $0.34$ & $0.34$ & $0.34$ & $0.32$ & $0.3$ & $0.27$ & $0.24$ & $0.21$ \\

$a_f^{7}$ & $0.18$ & $0.25$ & $0.29$ & $0.3$ & $0.3$ & $0.29$ & $0.27$ & $0.25$ & $0.22$ & $0.19$ \\

$a_f^{8}$ & $0.14$ & $0.21$ & $0.25$ & $0.27$ & $0.27$ & $0.26$ & $0.25$ & $0.23$ & $0.2$ & $0.17$ \\

$a_f^{9}$ & $0.1$ & $0.18$ & $0.22$ & $0.24$ & $0.25$ & $0.24$ & $0.23$ & $0.21$ & $0.19$ & $0.16$ \\

$a_f^{10}$ & $0.07$ & $0.15$ & $0.2$ & $0.22$ & $0.23$ & $0.23$ & $0.21$ & $0.2$ & $0.17$ & $0.15$ \\

\end{tabular}}
\caption{\label{tab:corr13} Correlations among the $a_f$ and $a_{\mathcal{F}_1}$ parameters.
}
\end{table}
\FloatBarrier

\begin{table}[h!]
\centering
{\footnotesize
\begin{tabular}{c|cccccccccc}
& $a_{P_1}^{1}$ & $a_{P_1}^{2}$ & $a_{P_1}^{3}$ & $a_{P_1}^{4}$ & $a_{P_1}^{5}$ & $a_{P_1}^{6}$ & $a_{P_1}^{7}$ & $a_{P_1}^{8}$ & $a_{P_1}^{9}$ & $a_{P_1}^{10}$  \\
\hline

$a_f^{1}$ & $0.61$ & $0.6$ & $0.59$ & $0.57$ & $0.55$ & $0.52$ & $0.48$ & $0.44$ & $0.39$ & $0.34$ \\

$a_f^{2}$ & $0.6$ & $0.61$ & $0.6$ & $0.59$ & $0.58$ & $0.56$ & $0.53$ & $0.5$ & $0.46$ & $0.42$ \\

$a_f^{3}$ & $0.54$ & $0.55$ & $0.55$ & $0.55$ & $0.55$ & $0.54$ & $0.52$ & $0.5$ & $0.47$ & $0.44$ \\

$a_f^{4}$ & $0.46$ & $0.48$ & $0.48$ & $0.49$ & $0.49$ & $0.49$ & $0.48$ & $0.47$ & $0.45$ & $0.43$ \\

$a_f^{5}$ & $0.39$ & $0.41$ & $0.42$ & $0.43$ & $0.44$ & $0.44$ & $0.44$ & $0.44$ & $0.43$ & $0.41$ \\

$a_f^{6}$ & $0.34$ & $0.35$ & $0.37$ & $0.38$ & $0.39$ & $0.4$ & $0.4$ & $0.4$ & $0.4$ & $0.39$ \\

$a_f^{7}$ & $0.29$ & $0.31$ & $0.33$ & $0.34$ & $0.36$ & $0.37$ & $0.37$ & $0.38$ & $0.38$ & $0.37$ \\

$a_f^{8}$ & $0.26$ & $0.28$ & $0.3$ & $0.31$ & $0.33$ & $0.34$ & $0.35$ & $0.36$ & $0.36$ & $0.36$ \\

$a_f^{9}$ & $0.23$ & $0.25$ & $0.27$ & $0.29$ & $0.3$ & $0.32$ & $0.33$ & $0.34$ & $0.34$ & $0.34$ \\

$a_f^{10}$ & $0.21$ & $0.23$ & $0.25$ & $0.26$ & $0.28$ & $0.3$ & $0.31$ & $0.32$ & $0.33$ & $0.33$ \\

\end{tabular}}
\caption{\label{tab:corr14} Correlations among the $a_f$ and $a_{P_1}$ parameters.
}
\end{table}
\FloatBarrier

\begin{table}[h!]
\centering
{\scriptsize
\begin{tabular}{c|cccccccccc}
& $b_f^{1}$ & $b_f^{2}$ & $b_f^{3}$ & $b_f^{4}$ & $b_f^{5}$ & $b_f^{6}$ & $b_f^{7}$ & $b_f^{8}$ & $b_f^{9}$ & $b_f^{10}$ \\
\hline

$a_f^{1}$ & $-0.99$ & $-0.95$ & $-0.82$ & $-0.68$ & $-0.56$ & $-0.46$ & $-0.39$ & $-0.34$ & $-0.29$ & $-0.26$ \\

$a_f^{2}$ & $-0.91$ & $-0.99$ & $-0.96$ & $-0.89$ & $-0.81$ & $-0.74$ & $-0.68$ & $-0.64$ & $-0.6$ & $-0.57$ \\

$a_f^{3}$ & $-0.75$ & $-0.93$ & $-0.99$ & $-0.98$ & $-0.94$ & $-0.9$ & $-0.87$ & $-0.84$ & $-0.81$ & $-0.79$ \\

$a_f^{4}$ & $-0.59$ & $-0.83$ & $-0.95$ & $-0.99$ & $-0.99$ & $-0.97$ & $-0.95$ & $-0.94$ & $-0.92$ & $-0.91$ \\

$a_f^{5}$ & $-0.45$ & $-0.73$ & $-0.9$ & $-0.97$ & $-0.99$ & $-0.99$ & $-0.99$ & $-0.98$ & $-0.97$ & $-0.96$ \\

$a_f^{6}$ & $-0.35$ & $-0.65$ & $-0.84$ & $-0.94$ & $-0.98$ & $-0.99$ & $-1.$ & $-0.99$ & $-0.99$ & $-0.98$ \\

$a_f^{7}$ & $-0.27$ & $-0.59$ & $-0.8$ & $-0.91$ & $-0.96$ & $-0.98$ & $-0.99$ & $-1.$ & $-1.$ & $-0.99$ \\

$a_f^{8}$ & $-0.21$ & $-0.54$ & $-0.76$ & $-0.88$ & $-0.94$ & $-0.97$ & $-0.99$ & $-0.99$ & $-1.$ & $-1.$ \\

$a_f^{9}$ & $-0.16$ & $-0.49$ & $-0.73$ & $-0.86$ & $-0.93$ & $-0.96$ & $-0.98$ & $-0.99$ & $-0.99$ & $-1.$ \\

$a_f^{10}$ & $-0.12$ & $-0.46$ & $-0.7$ & $-0.84$ & $-0.91$ & $-0.95$ & $-0.97$ & $-0.98$ & $-0.99$ & $-0.99$ \\

\end{tabular}}
\caption{\label{tab:corr15} Correlations among the $a_f$ and $b_f$ parameters.
}
\end{table}
\FloatBarrier

\begin{table}[h!]
\centering
{\scriptsize
\begin{tabular}{c|cccccccccc}
& $b_g^{1}$ & $b_g^{2}$ & $b_g^{3}$ & $b_g^{4}$ & $b_g^{5}$ & $b_g^{6}$ & $b_g^{7}$ & $b_g^{8}$ & $b_g^{9}$ & $b_g^{10}$  \\
\hline

$a_f^{1}$ & $-0.29$ & $-0.28$ & $-0.27$ & $-0.26$ & $-0.25$ & $-0.24$ & $-0.22$ & $-0.21$ & $-0.19$ & $-0.18$ \\

$a_f^{2}$ & $-0.35$ & $-0.34$ & $-0.34$ & $-0.32$ & $-0.31$ & $-0.3$ & $-0.28$ & $-0.26$ & $-0.24$ & $-0.23$ \\

$a_f^{3}$ & $-0.37$ & $-0.36$ & $-0.36$ & $-0.34$ & $-0.33$ & $-0.32$ & $-0.3$ & $-0.28$ & $-0.26$ & $-0.24$ \\

$a_f^{4}$ & $-0.36$ & $-0.36$ & $-0.35$ & $-0.34$ & $-0.33$ & $-0.31$ & $-0.29$ & $-0.28$ & $-0.26$ & $-0.24$ \\

$a_f^{5}$ & $-0.34$ & $-0.34$ & $-0.33$ & $-0.33$ & $-0.31$ & $-0.3$ & $-0.28$ & $-0.27$ & $-0.25$ & $-0.23$ \\

$a_f^{6}$ & $-0.33$ & $-0.32$ & $-0.32$ & $-0.31$ & $-0.3$ & $-0.29$ & $-0.27$ & $-0.26$ & $-0.24$ & $-0.23$ \\

$a_f^{7}$ & $-0.31$ & $-0.31$ & $-0.3$ & $-0.3$ & $-0.29$ & $-0.27$ & $-0.26$ & $-0.25$ & $-0.23$ & $-0.22$ \\

$a_f^{8}$ & $-0.3$ & $-0.3$ & $-0.29$ & $-0.28$ & $-0.28$ & $-0.26$ & $-0.25$ & $-0.24$ & $-0.22$ & $-0.21$ \\

$a_f^{9}$ & $-0.29$ & $-0.29$ & $-0.28$ & $-0.27$ & $-0.27$ & $-0.25$ & $-0.24$ & $-0.23$ & $-0.22$ & $-0.2$ \\

$a_f^{10}$ & $-0.28$ & $-0.28$ & $-0.27$ & $-0.27$ & $-0.26$ & $-0.25$ & $-0.23$ & $-0.22$ & $-0.21$ & $-0.2$ \\

\end{tabular}}
\caption{\label{tab:corr16} Correlations among the $a_f$ and $b_g$ parameters.
}
\end{table}
\FloatBarrier

\begin{table}[h!]
\centering
{\scriptsize
\begin{tabular}{c|cccccccccc}
& $b_{\mathcal{F}_1}^{1}$ & $b_{\mathcal{F}_1}^{2}$ & $b_{\mathcal{F}_1}^{3}$ & $b_{\mathcal{F}_1}^{4}$ & $b_{\mathcal{F}_1}^{5}$ & $b_{\mathcal{F}_1}^{6}$ & $b_{\mathcal{F}_1}^{7}$ & $b_{\mathcal{F}_1}^{8}$ & $b_{\mathcal{F}_1}^{9}$ & $b_{\mathcal{F}_1}^{10}$  \\
\hline

$a_f^{1}$ & $-0.69$ & $-0.67$ & $-0.62$ & $-0.56$ & $-0.51$ & $-0.47$ & $-0.43$ & $-0.4$ & $-0.37$ & $-0.34$ \\

$a_f^{2}$ & $-0.61$ & $-0.62$ & $-0.59$ & $-0.54$ & $-0.5$ & $-0.47$ & $-0.43$ & $-0.4$ & $-0.37$ & $-0.34$ \\

$a_f^{3}$ & $-0.48$ & $-0.51$ & $-0.5$ & $-0.48$ & $-0.45$ & $-0.42$ & $-0.39$ & $-0.37$ & $-0.34$ & $-0.31$ \\

$a_f^{4}$ & $-0.35$ & $-0.4$ & $-0.41$ & $-0.4$ & $-0.38$ & $-0.36$ & $-0.34$ & $-0.32$ & $-0.29$ & $-0.26$ \\

$a_f^{5}$ & $-0.25$ & $-0.31$ & $-0.33$ & $-0.33$ & $-0.33$ & $-0.31$ & $-0.29$ & $-0.27$ & $-0.25$ & $-0.23$ \\

$a_f^{6}$ & $-0.18$ & $-0.24$ & $-0.27$ & $-0.28$ & $-0.28$ & $-0.27$ & $-0.26$ & $-0.24$ & $-0.22$ & $-0.19$ \\

$a_f^{7}$ & $-0.12$ & $-0.19$ & $-0.22$ & $-0.24$ & $-0.24$ & $-0.24$ & $-0.23$ & $-0.21$ & $-0.19$ & $-0.17$ \\

$a_f^{8}$ & $-0.08$ & $-0.15$ & $-0.19$ & $-0.21$ & $-0.21$ & $-0.21$ & $-0.2$ & $-0.19$ & $-0.17$ & $-0.15$ \\

$a_f^{9}$ & $-0.04$ & $-0.11$ & $-0.16$ & $-0.18$ & $-0.19$ & $-0.19$ & $-0.18$ & $-0.17$ & $-0.15$ & $-0.14$ \\

$a_f^{10}$ & $-0.01$ & $-0.09$ & $-0.13$ & $-0.16$ & $-0.17$ & $-0.17$ & $-0.16$ & $-0.15$ & $-0.14$ & $-0.12$ \\

\end{tabular}}
\caption{\label{tab:corr17} Correlations among the $a_f$ and $b_{\mathcal{F}_1}$ parameters.
}
\end{table}
\FloatBarrier

\begin{table}[h!]
\centering
{\scriptsize
\begin{tabular}{c|cccccccccc}
& $b_{P_1}^{1}$ & $b_{P_1}^{2}$ & $b_{P_1}^{3}$ & $b_{P_1}^{4}$ & $b_{P_1}^{5}$ & $b_{P_1}^{6}$ & $b_{P_1}^{7}$ & $b_{P_1}^{8}$ & $b_{P_1}^{9}$ & $b_{P_1}^{10}$  \\
\hline

$a_f^{1}$ & $-0.61$ & $-0.6$ & $-0.59$ & $-0.58$ & $-0.56$ & $-0.54$ & $-0.51$ & $-0.48$ & $-0.46$ & $-0.43$ \\

$a_f^{2}$ & $-0.59$ & $-0.59$ & $-0.58$ & $-0.58$ & $-0.56$ & $-0.55$ & $-0.53$ & $-0.51$ & $-0.49$ & $-0.46$ \\

$a_f^{3}$ & $-0.51$ & $-0.52$ & $-0.52$ & $-0.52$ & $-0.51$ & $-0.5$ & $-0.49$ & $-0.48$ & $-0.46$ & $-0.44$ \\

$a_f^{4}$ & $-0.43$ & $-0.44$ & $-0.44$ & $-0.45$ & $-0.45$ & $-0.44$ & $-0.44$ & $-0.43$ & $-0.42$ & $-0.41$ \\

$a_f^{5}$ & $-0.36$ & $-0.37$ & $-0.38$ & $-0.38$ & $-0.38$ & $-0.38$ & $-0.38$ & $-0.38$ & $-0.37$ & $-0.37$ \\

$a_f^{6}$ & $-0.3$ & $-0.31$ & $-0.32$ & $-0.33$ & $-0.33$ & $-0.34$ & $-0.34$ & $-0.34$ & $-0.34$ & $-0.33$ \\

$a_f^{7}$ & $-0.25$ & $-0.27$ & $-0.28$ & $-0.29$ & $-0.3$ & $-0.3$ & $-0.3$ & $-0.31$ & $-0.31$ & $-0.3$ \\

$a_f^{8}$ & $-0.22$ & $-0.23$ & $-0.24$ & $-0.25$ & $-0.26$ & $-0.27$ & $-0.28$ & $-0.28$ & $-0.28$ & $-0.28$ \\

$a_f^{9}$ & $-0.19$ & $-0.2$ & $-0.22$ & $-0.23$ & $-0.24$ & $-0.25$ & $-0.25$ & $-0.26$ & $-0.26$ & $-0.26$ \\

$a_f^{10}$ & $-0.16$ & $-0.18$ & $-0.19$ & $-0.21$ & $-0.22$ & $-0.23$ & $-0.23$ & $-0.24$ & $-0.24$ & $-0.25$ \\

\end{tabular}}
\caption{\label{tab:corr18} Correlations among the $a_f$ and $b_{P_1}$ parameters.
}
\end{table}
\FloatBarrier

\begin{table}[h!]
\centering
{\footnotesize
\begin{tabular}{c|cccccccccc}
& $a_g^{1}$ & $a_g^{2}$ & $a_g^{3}$ & $a_g^{4}$ & $a_g^{5}$ & $a_g^{6}$ & $a_g^{7}$ & $a_g^{8}$ & $a_g^{9}$ & $a_g^{10}$ \\
\hline

$a_g^{1}$ & $1.$ & $1.$ & $0.98$ & $0.96$ & $0.92$ & $0.87$ & $0.82$ & $0.76$ & $0.69$ & $0.61$ \\

$a_g^{2}$ & $1.$ & $1.$ & $0.99$ & $0.98$ & $0.95$ & $0.92$ & $0.87$ & $0.82$ & $0.75$ & $0.69$ \\

$a_g^{3}$ & $0.98$ & $0.99$ & $1.$ & $0.99$ & $0.98$ & $0.95$ & $0.91$ & $0.87$ & $0.82$ & $0.76$ \\

$a_g^{4}$ & $0.96$ & $0.98$ & $0.99$ & $1.$ & $0.99$ & $0.98$ & $0.95$ & $0.92$ & $0.87$ & $0.82$ \\

$a_g^{5}$ & $0.92$ & $0.95$ & $0.98$ & $0.99$ & $1.$ & $0.99$ & $0.98$ & $0.95$ & $0.92$ & $0.87$ \\

$a_g^{6}$ & $0.87$ & $0.92$ & $0.95$ & $0.98$ & $0.99$ & $1.$ & $0.99$ & $0.98$ & $0.95$ & $0.92$ \\

$a_g^{7}$ & $0.82$ & $0.87$ & $0.91$ & $0.95$ & $0.98$ & $0.99$ & $1.$ & $0.99$ & $0.98$ & $0.96$ \\

$a_g^{8}$ & $0.76$ & $0.82$ & $0.87$ & $0.92$ & $0.95$ & $0.98$ & $0.99$ & $1.$ & $1.$ & $0.98$ \\

$a_g^{9}$ & $0.69$ & $0.75$ & $0.82$ & $0.87$ & $0.92$ & $0.95$ & $0.98$ & $1.$ & $1.$ & $1.$ \\

$a_g^{10}$ & $0.61$ & $0.69$ & $0.76$ & $0.82$ & $0.87$ & $0.92$ & $0.96$ & $0.98$ & $1.$ & $1.$ \\

\end{tabular}}
\caption{\label{tab:corr22} Correlations among the $a_g$ and $a_g$ parameters.
}
\end{table}
\FloatBarrier

\begin{table}[h!]
\centering
{\footnotesize
\begin{tabular}{c|cccccccccc}
& $a_{\mathcal{F}_1}^{1}$ & $a_{\mathcal{F}_1}^{2}$ & $a_{\mathcal{F}_1}^{3}$ & $a_{\mathcal{F}_1}^{4}$ & $a_{\mathcal{F}_1}^{5}$ & $a_{\mathcal{F}_1}^{6}$ & $a_{\mathcal{F}_1}^{7}$ & $a_{\mathcal{F}_1}^{8}$ & $a_{\mathcal{F}_1}^{9}$ & $a_{\mathcal{F}_1}^{10}$  \\
\hline

$a_g^{1}$ & $0.12$ & $0.18$ & $0.23$ & $0.27$ & $0.29$ & $0.3$ & $0.3$ & $0.3$ & $0.29$ & $0.28$ \\

$a_g^{2}$ & $0.11$ & $0.18$ & $0.23$ & $0.26$ & $0.28$ & $0.29$ & $0.3$ & $0.3$ & $0.29$ & $0.28$ \\

$a_g^{3}$ & $0.1$ & $0.17$ & $0.22$ & $0.25$ & $0.27$ & $0.29$ & $0.29$ & $0.29$ & $0.28$ & $0.27$ \\

$a_g^{4}$ & $0.09$ & $0.16$ & $0.21$ & $0.24$ & $0.26$ & $0.28$ & $0.28$ & $0.28$ & $0.27$ & $0.26$ \\

$a_g^{5}$ & $0.09$ & $0.15$ & $0.2$ & $0.23$ & $0.25$ & $0.26$ & $0.27$ & $0.26$ & $0.26$ & $0.25$ \\

$a_g^{6}$ & $0.07$ & $0.14$ & $0.18$ & $0.21$ & $0.23$ & $0.24$ & $0.25$ & $0.25$ & $0.24$ & $0.23$ \\

$a_g^{7}$ & $0.06$ & $0.12$ & $0.17$ & $0.19$ & $0.21$ & $0.22$ & $0.23$ & $0.23$ & $0.22$ & $0.21$ \\

$a_g^{8}$ & $0.05$ & $0.1$ & $0.15$ & $0.17$ & $0.19$ & $0.2$ & $0.2$ & $0.2$ & $0.2$ & $0.19$ \\

$a_g^{9}$ & $0.04$ & $0.09$ & $0.12$ & $0.15$ & $0.16$ & $0.17$ & $0.18$ & $0.18$ & $0.17$ & $0.16$ \\

$a_g^{10}$ & $0.02$ & $0.07$ & $0.1$ & $0.12$ & $0.14$ & $0.15$ & $0.15$ & $0.15$ & $0.14$ & $0.14$ \\

\end{tabular}}
\caption{\label{tab:corr23} Correlations among the $a_g$ and $a_{\mathcal{F}_1}$ parameters.
}
\end{table}
\FloatBarrier

\begin{table}[h!]
\centering
{\footnotesize
\begin{tabular}{c|cccccccccc}
& $a_{P_1}^{1}$ & $a_{P_1}^{2}$ & $a_{P_1}^{3}$ & $a_{P_1}^{4}$ & $a_{P_1}^{5}$ & $a_{P_1}^{6}$ & $a_{P_1}^{7}$ & $a_{P_1}^{8}$ & $a_{P_1}^{9}$ & $a_{P_1}^{10}$  \\
\hline

$a_g^{1}$ & $0.41$ & $0.41$ & $0.41$ & $0.4$ & $0.39$ & $0.38$ & $0.36$ & $0.33$ & $0.31$ & $0.27$ \\

$a_g^{2}$ & $0.41$ & $0.41$ & $0.41$ & $0.4$ & $0.39$ & $0.38$ & $0.36$ & $0.33$ & $0.31$ & $0.28$ \\

$a_g^{3}$ & $0.4$ & $0.4$ & $0.4$ & $0.39$ & $0.38$ & $0.37$ & $0.35$ & $0.33$ & $0.3$ & $0.27$ \\

$a_g^{4}$ & $0.39$ & $0.39$ & $0.39$ & $0.38$ & $0.38$ & $0.36$ & $0.34$ & $0.32$ & $0.3$ & $0.27$ \\

$a_g^{5}$ & $0.37$ & $0.38$ & $0.37$ & $0.37$ & $0.36$ & $0.35$ & $0.33$ & $0.31$ & $0.29$ & $0.26$ \\

$a_g^{6}$ & $0.35$ & $0.36$ & $0.35$ & $0.35$ & $0.34$ & $0.33$ & $0.32$ & $0.3$ & $0.28$ & $0.25$ \\

$a_g^{7}$ & $0.33$ & $0.33$ & $0.33$ & $0.33$ & $0.32$ & $0.31$ & $0.3$ & $0.28$ & $0.26$ & $0.24$ \\

$a_g^{8}$ & $0.3$ & $0.3$ & $0.3$ & $0.3$ & $0.3$ & $0.29$ & $0.28$ & $0.26$ & $0.25$ & $0.23$ \\

$a_g^{9}$ & $0.27$ & $0.28$ & $0.28$ & $0.27$ & $0.27$ & $0.26$ & $0.26$ & $0.24$ & $0.23$ & $0.21$ \\

$a_g^{10}$ & $0.24$ & $0.24$ & $0.24$ & $0.24$ & $0.24$ & $0.24$ & $0.23$ & $0.22$ & $0.21$ & $0.19$ \\

\end{tabular}}
\caption{\label{tab:corr24} Correlations among the $a_g$ and $a_{P_1}$ parameters.
}
\end{table}
\FloatBarrier

\begin{table}[h!]
\centering
{\scriptsize
\begin{tabular}{c|cccccccccc}
& $b_f^{1}$ & $b_f^{2}$ & $b_f^{3}$ & $b_f^{4}$ & $b_f^{5}$ & $b_f^{6}$ & $b_f^{7}$ & $b_f^{8}$ & $b_f^{9}$ & $b_f^{10}$ \\
\hline

$a_g^{1}$ & $-0.26$ & $-0.33$ & $-0.36$ & $-0.36$ & $-0.35$ & $-0.34$ & $-0.32$ & $-0.31$ & $-0.31$ & $-0.3$ \\

$a_g^{2}$ & $-0.26$ & $-0.33$ & $-0.35$ & $-0.35$ & $-0.34$ & $-0.33$ & $-0.32$ & $-0.31$ & $-0.3$ & $-0.3$ \\

$a_g^{3}$ & $-0.25$ & $-0.32$ & $-0.34$ & $-0.34$ & $-0.34$ & $-0.32$ & $-0.31$ & $-0.3$ & $-0.3$ & $-0.29$ \\

$a_g^{4}$ & $-0.23$ & $-0.3$ & $-0.33$ & $-0.33$ & $-0.32$ & $-0.31$ & $-0.3$ & $-0.29$ & $-0.29$ & $-0.28$ \\

$a_g^{5}$ & $-0.22$ & $-0.28$ & $-0.31$ & $-0.31$ & $-0.31$ & $-0.3$ & $-0.29$ & $-0.28$ & $-0.27$ & $-0.27$ \\

$a_g^{6}$ & $-0.2$ & $-0.26$ & $-0.29$ & $-0.29$ & $-0.29$ & $-0.28$ & $-0.27$ & $-0.26$ & $-0.26$ & $-0.25$ \\

$a_g^{7}$ & $-0.18$ & $-0.23$ & $-0.26$ & $-0.27$ & $-0.26$ & $-0.26$ & $-0.25$ & $-0.24$ & $-0.24$ & $-0.23$ \\

$a_g^{8}$ & $-0.15$ & $-0.21$ & $-0.23$ & $-0.24$ & $-0.24$ & $-0.23$ & $-0.23$ & $-0.22$ & $-0.22$ & $-0.21$ \\

$a_g^{9}$ & $-0.13$ & $-0.18$ & $-0.2$ & $-0.21$ & $-0.21$ & $-0.2$ & $-0.2$ & $-0.2$ & $-0.19$ & $-0.19$ \\

$a_g^{10}$ & $-0.1$ & $-0.15$ & $-0.17$ & $-0.18$ & $-0.18$ & $-0.18$ & $-0.17$ & $-0.17$ & $-0.17$ & $-0.17$ \\

\end{tabular}}
\caption{\label{tab:corr25} Correlations among the $a_g$ and $b_f$ parameters.
}
\end{table}
\FloatBarrier

\begin{table}[h!]
\centering
{\scriptsize
\begin{tabular}{c|cccccccccc}
& $b_g^{1}$ & $b_g^{2}$ & $b_g^{3}$ & $b_g^{4}$ & $b_g^{5}$ & $b_g^{6}$ & $b_g^{7}$ & $b_g^{8}$ & $b_g^{9}$ & $b_g^{10}$  \\
\hline

$a_g^{1}$ & $-0.99$ & $-0.98$ & $-0.97$ & $-0.95$ & $-0.92$ & $-0.88$ & $-0.84$ & $-0.79$ & $-0.75$ & $-0.7$ \\

$a_g^{2}$ & $-0.98$ & $-0.99$ & $-0.98$ & $-0.97$ & $-0.94$ & $-0.92$ & $-0.88$ & $-0.84$ & $-0.8$ & $-0.76$ \\

$a_g^{3}$ & $-0.96$ & $-0.98$ & $-0.98$ & $-0.98$ & $-0.97$ & $-0.94$ & $-0.92$ & $-0.89$ & $-0.85$ & $-0.82$ \\

$a_g^{4}$ & $-0.93$ & $-0.96$ & $-0.97$ & $-0.98$ & $-0.98$ & $-0.96$ & $-0.95$ & $-0.92$ & $-0.9$ & $-0.87$ \\

$a_g^{5}$ & $-0.89$ & $-0.93$ & $-0.95$ & $-0.97$ & $-0.97$ & $-0.97$ & $-0.96$ & $-0.95$ & $-0.93$ & $-0.91$ \\

$a_g^{6}$ & $-0.84$ & $-0.89$ & $-0.92$ & $-0.95$ & $-0.96$ & $-0.97$ & $-0.97$ & $-0.96$ & $-0.95$ & $-0.94$ \\

$a_g^{7}$ & $-0.78$ & $-0.84$ & $-0.88$ & $-0.91$ & $-0.94$ & $-0.96$ & $-0.96$ & $-0.97$ & $-0.96$ & $-0.95$ \\

$a_g^{8}$ & $-0.72$ & $-0.78$ & $-0.83$ & $-0.87$ & $-0.91$ & $-0.93$ & $-0.95$ & $-0.96$ & $-0.96$ & $-0.96$ \\

$a_g^{9}$ & $-0.64$ & $-0.71$ & $-0.77$ & $-0.82$ & $-0.87$ & $-0.9$ & $-0.92$ & $-0.94$ & $-0.95$ & $-0.96$ \\

$a_g^{10}$ & $-0.57$ & $-0.64$ & $-0.71$ & $-0.77$ & $-0.82$ & $-0.86$ & $-0.89$ & $-0.91$ & $-0.93$ & $-0.94$ \\

\end{tabular}}
\caption{\label{tab:corr26} Correlations among the $a_g$ and $b_g$ parameters.
}
\end{table}
\FloatBarrier

\begin{table}[h!]
\centering
{\scriptsize
\begin{tabular}{c|cccccccccc}
& $b_{\mathcal{F}_1}^{1}$ & $b_{\mathcal{F}_1}^{2}$ & $b_{\mathcal{F}_1}^{3}$ & $b_{\mathcal{F}_1}^{4}$ & $b_{\mathcal{F}_1}^{5}$ & $b_{\mathcal{F}_1}^{6}$ & $b_{\mathcal{F}_1}^{7}$ & $b_{\mathcal{F}_1}^{8}$ & $b_{\mathcal{F}_1}^{9}$ & $b_{\mathcal{F}_1}^{10}$  \\
\hline

$a_g^{1}$ & $-0.11$ & $-0.17$ & $-0.22$ & $-0.25$ & $-0.27$ & $-0.28$ & $-0.29$ & $-0.29$ & $-0.29$ & $-0.28$ \\

$a_g^{2}$ & $-0.1$ & $-0.17$ & $-0.22$ & $-0.25$ & $-0.27$ & $-0.28$ & $-0.28$ & $-0.29$ & $-0.28$ & $-0.28$ \\

$a_g^{3}$ & $-0.1$ & $-0.16$ & $-0.21$ & $-0.24$ & $-0.26$ & $-0.27$ & $-0.28$ & $-0.28$ & $-0.28$ & $-0.27$ \\

$a_g^{4}$ & $-0.09$ & $-0.15$ & $-0.2$ & $-0.23$ & $-0.25$ & $-0.26$ & $-0.27$ & $-0.27$ & $-0.27$ & $-0.26$ \\

$a_g^{5}$ & $-0.08$ & $-0.14$ & $-0.19$ & $-0.22$ & $-0.24$ & $-0.25$ & $-0.25$ & $-0.25$ & $-0.25$ & $-0.25$ \\

$a_g^{6}$ & $-0.07$ & $-0.13$ & $-0.17$ & $-0.2$ & $-0.22$ & $-0.23$ & $-0.24$ & $-0.24$ & $-0.23$ & $-0.23$ \\

$a_g^{7}$ & $-0.06$ & $-0.11$ & $-0.16$ & $-0.18$ & $-0.2$ & $-0.21$ & $-0.22$ & $-0.22$ & $-0.21$ & $-0.21$ \\

$a_g^{8}$ & $-0.04$ & $-0.1$ & $-0.14$ & $-0.16$ & $-0.18$ & $-0.19$ & $-0.19$ & $-0.19$ & $-0.19$ & $-0.19$ \\

$a_g^{9}$ & $-0.03$ & $-0.08$ & $-0.12$ & $-0.14$ & $-0.15$ & $-0.16$ & $-0.17$ & $-0.17$ & $-0.17$ & $-0.16$ \\

$a_g^{10}$ & $-0.02$ & $-0.06$ & $-0.09$ & $-0.11$ & $-0.13$ & $-0.14$ & $-0.14$ & $-0.14$ & $-0.14$ & $-0.14$ \\

\end{tabular}}
\caption{\label{tab:corr27} Correlations among the $a_g$ and $b_{\mathcal{F}_1}$ parameters.
}
\end{table}
\FloatBarrier

\begin{table}[h!]
\centering
{\scriptsize
\begin{tabular}{c|cccccccccc}
& $b_{P_1}^{1}$ & $b_{P_1}^{2}$ & $b_{P_1}^{3}$ & $b_{P_1}^{4}$ & $b_{P_1}^{5}$ & $b_{P_1}^{6}$ & $b_{P_1}^{7}$ & $b_{P_1}^{8}$ & $b_{P_1}^{9}$ & $b_{P_1}^{10}$  \\
\hline

$a_g^{1}$ & $-0.4$ & $-0.4$ & $-0.39$ & $-0.39$ & $-0.38$ & $-0.36$ & $-0.35$ & $-0.34$ & $-0.32$ & $-0.3$ \\

$a_g^{2}$ & $-0.4$ & $-0.4$ & $-0.39$ & $-0.38$ & $-0.38$ & $-0.36$ & $-0.35$ & $-0.33$ & $-0.32$ & $-0.3$ \\

$a_g^{3}$ & $-0.39$ & $-0.39$ & $-0.38$ & $-0.38$ & $-0.37$ & $-0.36$ & $-0.34$ & $-0.33$ & $-0.31$ & $-0.3$ \\

$a_g^{4}$ & $-0.38$ & $-0.38$ & $-0.37$ & $-0.37$ & $-0.36$ & $-0.35$ & $-0.34$ & $-0.32$ & $-0.31$ & $-0.29$ \\

$a_g^{5}$ & $-0.36$ & $-0.36$ & $-0.36$ & $-0.35$ & $-0.34$ & $-0.33$ & $-0.32$ & $-0.31$ & $-0.3$ & $-0.28$ \\

$a_g^{6}$ & $-0.34$ & $-0.34$ & $-0.34$ & $-0.33$ & $-0.33$ & $-0.32$ & $-0.31$ & $-0.29$ & $-0.28$ & $-0.27$ \\

$a_g^{7}$ & $-0.31$ & $-0.31$ & $-0.31$ & $-0.31$ & $-0.3$ & $-0.29$ & $-0.29$ & $-0.27$ & $-0.26$ & $-0.25$ \\

$a_g^{8}$ & $-0.29$ & $-0.29$ & $-0.29$ & $-0.28$ & $-0.28$ & $-0.27$ & $-0.26$ & $-0.25$ & $-0.24$ & $-0.23$ \\

$a_g^{9}$ & $-0.26$ & $-0.26$ & $-0.26$ & $-0.25$ & $-0.25$ & $-0.24$ & $-0.24$ & $-0.23$ & $-0.22$ & $-0.21$ \\

$a_g^{10}$ & $-0.22$ & $-0.22$ & $-0.22$ & $-0.22$ & $-0.22$ & $-0.21$ & $-0.21$ & $-0.2$ & $-0.19$ & $-0.19$ \\

\end{tabular}}
\caption{\label{tab:corr28} Correlations among the $a_g$ and $b_{P_1}$ parameters.
}
\end{table}
\FloatBarrier

\begin{table}[h!]
\centering
{\footnotesize
\begin{tabular}{c|cccccccccc}
& $a_{\mathcal{F}_1}^{1}$ & $a_{\mathcal{F}_1}^{2}$ & $a_{\mathcal{F}_1}^{3}$ & $a_{\mathcal{F}_1}^{4}$ & $a_{\mathcal{F}_1}^{5}$ & $a_{\mathcal{F}_1}^{6}$ & $a_{\mathcal{F}_1}^{7}$ & $a_{\mathcal{F}_1}^{8}$ & $a_{\mathcal{F}_1}^{9}$ & $a_{\mathcal{F}_1}^{10}$  \\
\hline

$a_{\mathcal{F}_1}^{1}$ & $1.$ & $0.95$ & $0.85$ & $0.74$ & $0.64$ & $0.55$ & $0.48$ & $0.41$ & $0.36$ & $0.31$ \\

$a_{\mathcal{F}_1}^{2}$ & $0.95$ & $1.$ & $0.97$ & $0.9$ & $0.83$ & $0.76$ & $0.69$ & $0.61$ & $0.54$ & $0.47$ \\

$a_{\mathcal{F}_1}^{3}$ & $0.85$ & $0.97$ & $1.$ & $0.98$ & $0.94$ & $0.89$ & $0.83$ & $0.76$ & $0.68$ & $0.6$ \\

$a_{\mathcal{F}_1}^{4}$ & $0.74$ & $0.9$ & $0.98$ & $1.$ & $0.99$ & $0.96$ & $0.91$ & $0.85$ & $0.79$ & $0.71$ \\

$a_{\mathcal{F}_1}^{5}$ & $0.64$ & $0.83$ & $0.94$ & $0.99$ & $1.$ & $0.99$ & $0.96$ & $0.92$ & $0.86$ & $0.79$ \\

$a_{\mathcal{F}_1}^{6}$ & $0.55$ & $0.76$ & $0.89$ & $0.96$ & $0.99$ & $1.$ & $0.99$ & $0.96$ & $0.92$ & $0.87$ \\

$a_{\mathcal{F}_1}^{7}$ & $0.48$ & $0.69$ & $0.83$ & $0.91$ & $0.96$ & $0.99$ & $1.$ & $0.99$ & $0.97$ & $0.92$ \\

$a_{\mathcal{F}_1}^{8}$ & $0.41$ & $0.61$ & $0.76$ & $0.85$ & $0.92$ & $0.96$ & $0.99$ & $1.$ & $0.99$ & $0.97$ \\

$a_{\mathcal{F}_1}^{9}$ & $0.36$ & $0.54$ & $0.68$ & $0.79$ & $0.86$ & $0.92$ & $0.97$ & $0.99$ & $1.$ & $0.99$ \\

$a_{\mathcal{F}_1}^{10}$ & $0.31$ & $0.47$ & $0.6$ & $0.71$ & $0.79$ & $0.87$ & $0.92$ & $0.97$ & $0.99$ & $1.$ \\

\end{tabular}}
\caption{\label{tab:corr33} Correlations among the $a_{\mathcal{F}_1}$ and $a_{\mathcal{F}_1}$ parameters.
}
\end{table}
\FloatBarrier

\begin{table}[h!]
\centering
{\footnotesize
\begin{tabular}{c|cccccccccc}
& $a_{P_1}^{1}$ & $a_{P_1}^{2}$ & $a_{P_1}^{3}$ & $a_{P_1}^{4}$ & $a_{P_1}^{5}$ & $a_{P_1}^{6}$ & $a_{P_1}^{7}$ & $a_{P_1}^{8}$ & $a_{P_1}^{9}$ & $a_{P_1}^{10}$  \\
\hline

$a_{\mathcal{F}_1}^{1}$ & $0.51$ & $0.5$ & $0.48$ & $0.46$ & $0.42$ & $0.39$ & $0.34$ & $0.29$ & $0.24$ & $0.18$ \\

$a_{\mathcal{F}_1}^{2}$ & $0.65$ & $0.64$ & $0.63$ & $0.61$ & $0.58$ & $0.55$ & $0.5$ & $0.45$ & $0.4$ & $0.34$ \\

$a_{\mathcal{F}_1}^{3}$ & $0.72$ & $0.72$ & $0.72$ & $0.71$ & $0.68$ & $0.65$ & $0.62$ & $0.57$ & $0.51$ & $0.45$ \\

$a_{\mathcal{F}_1}^{4}$ & $0.76$ & $0.76$ & $0.77$ & $0.76$ & $0.74$ & $0.72$ & $0.69$ & $0.64$ & $0.59$ & $0.53$ \\

$a_{\mathcal{F}_1}^{5}$ & $0.77$ & $0.78$ & $0.79$ & $0.79$ & $0.78$ & $0.76$ & $0.73$ & $0.69$ & $0.64$ & $0.59$ \\

$a_{\mathcal{F}_1}^{6}$ & $0.77$ & $0.78$ & $0.8$ & $0.8$ & $0.79$ & $0.78$ & $0.76$ & $0.72$ & $0.68$ & $0.63$ \\

$a_{\mathcal{F}_1}^{7}$ & $0.75$ & $0.77$ & $0.79$ & $0.8$ & $0.8$ & $0.79$ & $0.77$ & $0.74$ & $0.7$ & $0.65$ \\

$a_{\mathcal{F}_1}^{8}$ & $0.73$ & $0.75$ & $0.77$ & $0.78$ & $0.79$ & $0.78$ & $0.77$ & $0.74$ & $0.71$ & $0.66$ \\

$a_{\mathcal{F}_1}^{9}$ & $0.7$ & $0.72$ & $0.74$ & $0.76$ & $0.76$ & $0.76$ & $0.75$ & $0.73$ & $0.7$ & $0.66$ \\

$a_{\mathcal{F}_1}^{10}$ & $0.66$ & $0.68$ & $0.7$ & $0.72$ & $0.73$ & $0.73$ & $0.72$ & $0.71$ & $0.68$ & $0.64$ \\

\end{tabular}}
\caption{\label{tab:corr34} Correlations among the $a_{\mathcal{F}_1}$ and $a_{P_1}$ parameters.
}
\end{table}
\FloatBarrier

\begin{table}[h!]
\centering
{\scriptsize
\begin{tabular}{c|cccccccccc}
& $b_f^{1}$ & $b_f^{2}$ & $b_f^{3}$ & $b_f^{4}$ & $b_f^{5}$ & $b_f^{6}$ & $b_f^{7}$ & $b_f^{8}$ & $b_f^{9}$ & $b_f^{10}$ \\
\hline

$a_{\mathcal{F}_1}^{1}$ & $-0.71$ & $-0.66$ & $-0.55$ & $-0.44$ & $-0.35$ & $-0.28$ & $-0.22$ & $-0.18$ & $-0.15$ & $-0.12$ \\

$a_{\mathcal{F}_1}^{2}$ & $-0.71$ & $-0.69$ & $-0.6$ & $-0.5$ & $-0.42$ & $-0.35$ & $-0.3$ & $-0.26$ & $-0.23$ & $-0.2$ \\

$a_{\mathcal{F}_1}^{3}$ & $-0.66$ & $-0.66$ & $-0.6$ & $-0.51$ & $-0.44$ & $-0.38$ & $-0.34$ & $-0.3$ & $-0.28$ & $-0.25$ \\

$a_{\mathcal{F}_1}^{4}$ & $-0.6$ & $-0.61$ & $-0.57$ & $-0.5$ & $-0.44$ & $-0.39$ & $-0.35$ & $-0.32$ & $-0.3$ & $-0.28$ \\

$a_{\mathcal{F}_1}^{5}$ & $-0.54$ & $-0.56$ & $-0.53$ & $-0.48$ & $-0.43$ & $-0.38$ & $-0.35$ & $-0.32$ & $-0.3$ & $-0.28$ \\

$a_{\mathcal{F}_1}^{6}$ & $-0.48$ & $-0.52$ & $-0.49$ & $-0.45$ & $-0.4$ & $-0.37$ & $-0.34$ & $-0.31$ & $-0.29$ & $-0.28$ \\

$a_{\mathcal{F}_1}^{7}$ & $-0.44$ & $-0.47$ & $-0.45$ & $-0.41$ & $-0.38$ & $-0.34$ & $-0.32$ & $-0.29$ & $-0.28$ & $-0.26$ \\

$a_{\mathcal{F}_1}^{8}$ & $-0.39$ & $-0.43$ & $-0.41$ & $-0.38$ & $-0.34$ & $-0.31$ & $-0.29$ & $-0.27$ & $-0.25$ & $-0.24$ \\

$a_{\mathcal{F}_1}^{9}$ & $-0.35$ & $-0.38$ & $-0.37$ & $-0.34$ & $-0.31$ & $-0.28$ & $-0.26$ & $-0.24$ & $-0.23$ & $-0.21$ \\

$a_{\mathcal{F}_1}^{10}$ & $-0.31$ & $-0.34$ & $-0.32$ & $-0.29$ & $-0.27$ & $-0.24$ & $-0.22$ & $-0.21$ & $-0.19$ & $-0.18$ \\

\end{tabular}}
\caption{\label{tab:corr35} Correlations among the $a_{\mathcal{F}_1}$ and $b_f$ parameters.
}
\end{table}
\FloatBarrier

\begin{table}[h!]
\centering
{\scriptsize
\begin{tabular}{c|cccccccccc}
& $b_g^{1}$ & $b_g^{2}$ & $b_g^{3}$ & $b_g^{4}$ & $b_g^{5}$ & $b_g^{6}$ & $b_g^{7}$ & $b_g^{8}$ & $b_g^{9}$ & $b_g^{10}$  \\
\hline

$a_{\mathcal{F}_1}^{1}$ & $-0.14$ & $-0.14$ & $-0.14$ & $-0.13$ & $-0.13$ & $-0.12$ & $-0.11$ & $-0.1$ & $-0.1$ & $-0.09$ \\

$a_{\mathcal{F}_1}^{2}$ & $-0.22$ & $-0.22$ & $-0.21$ & $-0.21$ & $-0.2$ & $-0.19$ & $-0.19$ & $-0.18$ & $-0.17$ & $-0.16$ \\

$a_{\mathcal{F}_1}^{3}$ & $-0.27$ & $-0.27$ & $-0.27$ & $-0.26$ & $-0.26$ & $-0.25$ & $-0.24$ & $-0.23$ & $-0.21$ & $-0.2$ \\

$a_{\mathcal{F}_1}^{4}$ & $-0.3$ & $-0.3$ & $-0.3$ & $-0.3$ & $-0.29$ & $-0.28$ & $-0.27$ & $-0.26$ & $-0.25$ & $-0.23$ \\

$a_{\mathcal{F}_1}^{5}$ & $-0.32$ & $-0.33$ & $-0.32$ & $-0.32$ & $-0.31$ & $-0.3$ & $-0.29$ & $-0.28$ & $-0.27$ & $-0.25$ \\

$a_{\mathcal{F}_1}^{6}$ & $-0.34$ & $-0.34$ & $-0.34$ & $-0.33$ & $-0.32$ & $-0.31$ & $-0.3$ & $-0.29$ & $-0.28$ & $-0.26$ \\

$a_{\mathcal{F}_1}^{7}$ & $-0.34$ & $-0.34$ & $-0.34$ & $-0.34$ & $-0.33$ & $-0.32$ & $-0.31$ & $-0.29$ & $-0.28$ & $-0.26$ \\

$a_{\mathcal{F}_1}^{8}$ & $-0.34$ & $-0.34$ & $-0.34$ & $-0.33$ & $-0.33$ & $-0.32$ & $-0.3$ & $-0.29$ & $-0.28$ & $-0.26$ \\

$a_{\mathcal{F}_1}^{9}$ & $-0.33$ & $-0.33$ & $-0.33$ & $-0.33$ & $-0.32$ & $-0.31$ & $-0.3$ & $-0.28$ & $-0.27$ & $-0.25$ \\

$a_{\mathcal{F}_1}^{10}$ & $-0.32$ & $-0.32$ & $-0.32$ & $-0.31$ & $-0.3$ & $-0.29$ & $-0.28$ & $-0.27$ & $-0.26$ & $-0.24$ \\

\end{tabular}}
\caption{\label{tab:corr36} Correlations among the $a_{\mathcal{F}_1}$ and $b_g$ parameters.
}
\end{table}
\FloatBarrier

\begin{table}[h!]
\centering
{\scriptsize
\begin{tabular}{c|cccccccccc}
& $b_{\mathcal{F}_1}^{1}$ & $b_{\mathcal{F}_1}^{2}$ & $b_{\mathcal{F}_1}^{3}$ & $b_{\mathcal{F}_1}^{4}$ & $b_{\mathcal{F}_1}^{5}$ & $b_{\mathcal{F}_1}^{6}$ & $b_{\mathcal{F}_1}^{7}$ & $b_{\mathcal{F}_1}^{8}$ & $b_{\mathcal{F}_1}^{9}$ & $b_{\mathcal{F}_1}^{10}$  \\
\hline

$a_{\mathcal{F}_1}^{1}$ & $-0.99$ & $-0.94$ & $-0.84$ & $-0.74$ & $-0.66$ & $-0.59$ & $-0.53$ & $-0.48$ & $-0.43$ & $-0.39$ \\

$a_{\mathcal{F}_1}^{2}$ & $-0.95$ & $-0.99$ & $-0.96$ & $-0.9$ & $-0.84$ & $-0.79$ & $-0.73$ & $-0.67$ & $-0.62$ & $-0.56$ \\

$a_{\mathcal{F}_1}^{3}$ & $-0.85$ & $-0.96$ & $-0.99$ & $-0.98$ & $-0.94$ & $-0.9$ & $-0.86$ & $-0.81$ & $-0.75$ & $-0.69$ \\

$a_{\mathcal{F}_1}^{4}$ & $-0.74$ & $-0.9$ & $-0.97$ & $-0.99$ & $-0.99$ & $-0.96$ & $-0.93$ & $-0.89$ & $-0.84$ & $-0.79$ \\

$a_{\mathcal{F}_1}^{5}$ & $-0.65$ & $-0.83$ & $-0.93$ & $-0.98$ & $-0.99$ & $-0.99$ & $-0.97$ & $-0.95$ & $-0.91$ & $-0.86$ \\

$a_{\mathcal{F}_1}^{6}$ & $-0.56$ & $-0.76$ & $-0.88$ & $-0.94$ & $-0.98$ & $-0.99$ & $-0.99$ & $-0.98$ & $-0.95$ & $-0.92$ \\

$a_{\mathcal{F}_1}^{7}$ & $-0.49$ & $-0.69$ & $-0.82$ & $-0.9$ & $-0.95$ & $-0.98$ & $-0.99$ & $-0.99$ & $-0.98$ & $-0.96$ \\

$a_{\mathcal{F}_1}^{8}$ & $-0.43$ & $-0.62$ & $-0.75$ & $-0.84$ & $-0.9$ & $-0.95$ & $-0.97$ & $-0.99$ & $-0.99$ & $-0.99$ \\

$a_{\mathcal{F}_1}^{9}$ & $-0.37$ & $-0.55$ & $-0.68$ & $-0.77$ & $-0.84$ & $-0.9$ & $-0.94$ & $-0.97$ & $-0.99$ & $-0.99$ \\

$a_{\mathcal{F}_1}^{10}$ & $-0.32$ & $-0.48$ & $-0.6$ & $-0.7$ & $-0.77$ & $-0.84$ & $-0.89$ & $-0.94$ & $-0.97$ & $-0.99$ \\

\end{tabular}}
\caption{\label{tab:corr37} Correlations among the $a_{\mathcal{F}_1}$ and $b_{\mathcal{F}_1}$ parameters.
}
\end{table}
\FloatBarrier

\begin{table}[h!]
\centering
{\scriptsize
\begin{tabular}{c|cccccccccc}
& $b_{P_1}^{1}$ & $b_{P_1}^{2}$ & $b_{P_1}^{3}$ & $b_{P_1}^{4}$ & $b_{P_1}^{5}$ & $b_{P_1}^{6}$ & $b_{P_1}^{7}$ & $b_{P_1}^{8}$ & $b_{P_1}^{9}$ & $b_{P_1}^{10}$  \\
\hline

$a_{\mathcal{F}_1}^{1}$ & $-0.57$ & $-0.57$ & $-0.56$ & $-0.54$ & $-0.52$ & $-0.5$ & $-0.48$ & $-0.45$ & $-0.43$ & $-0.4$ \\

$a_{\mathcal{F}_1}^{2}$ & $-0.72$ & $-0.72$ & $-0.72$ & $-0.71$ & $-0.7$ & $-0.68$ & $-0.66$ & $-0.64$ & $-0.61$ & $-0.58$ \\

$a_{\mathcal{F}_1}^{3}$ & $-0.79$ & $-0.81$ & $-0.81$ & $-0.81$ & $-0.8$ & $-0.79$ & $-0.78$ & $-0.76$ & $-0.74$ & $-0.71$ \\

$a_{\mathcal{F}_1}^{4}$ & $-0.83$ & $-0.85$ & $-0.86$ & $-0.86$ & $-0.86$ & $-0.86$ & $-0.85$ & $-0.83$ & $-0.81$ & $-0.79$ \\

$a_{\mathcal{F}_1}^{5}$ & $-0.84$ & $-0.86$ & $-0.88$ & $-0.89$ & $-0.89$ & $-0.89$ & $-0.89$ & $-0.87$ & $-0.86$ & $-0.84$ \\

$a_{\mathcal{F}_1}^{6}$ & $-0.83$ & $-0.86$ & $-0.88$ & $-0.9$ & $-0.91$ & $-0.91$ & $-0.91$ & $-0.9$ & $-0.88$ & $-0.87$ \\

$a_{\mathcal{F}_1}^{7}$ & $-0.82$ & $-0.85$ & $-0.87$ & $-0.89$ & $-0.9$ & $-0.91$ & $-0.91$ & $-0.9$ & $-0.89$ & $-0.88$ \\

$a_{\mathcal{F}_1}^{8}$ & $-0.79$ & $-0.82$ & $-0.85$ & $-0.87$ & $-0.89$ & $-0.89$ & $-0.9$ & $-0.89$ & $-0.89$ & $-0.88$ \\

$a_{\mathcal{F}_1}^{9}$ & $-0.75$ & $-0.79$ & $-0.82$ & $-0.84$ & $-0.86$ & $-0.87$ & $-0.87$ & $-0.87$ & $-0.87$ & $-0.86$ \\

$a_{\mathcal{F}_1}^{10}$ & $-0.71$ & $-0.74$ & $-0.77$ & $-0.79$ & $-0.81$ & $-0.82$ & $-0.83$ & $-0.83$ & $-0.83$ & $-0.82$ \\

\end{tabular}}
\caption{\label{tab:corr38} Correlations among the $a_{\mathcal{F}_1}$ and $b_{P_1}$ parameters.
}
\end{table}
\FloatBarrier

\begin{table}[h!]
\centering
{\footnotesize
\begin{tabular}{c|cccccccccc}
& $a_{P_1}^{1}$ & $a_{P_1}^{2}$ & $a_{P_1}^{3}$ & $a_{P_1}^{4}$ & $a_{P_1}^{5}$ & $a_{P_1}^{6}$ & $a_{P_1}^{7}$ & $a_{P_1}^{8}$ & $a_{P_1}^{9}$ & $a_{P_1}^{10}$  \\
\hline

$a_{P_1}^{1}$ & $1.$ & $1.$ & $0.99$ & $0.97$ & $0.94$ & $0.89$ & $0.84$ & $0.78$ & $0.71$ & $0.63$ \\

$a_{P_1}^{2}$ & $1.$ & $1.$ & $1.$ & $0.98$ & $0.96$ & $0.93$ & $0.89$ & $0.83$ & $0.77$ & $0.7$ \\

$a_{P_1}^{3}$ & $0.99$ & $1.$ & $1.$ & $1.$ & $0.98$ & $0.96$ & $0.92$ & $0.88$ & $0.82$ & $0.76$ \\

$a_{P_1}^{4}$ & $0.97$ & $0.98$ & $1.$ & $1.$ & $1.$ & $0.98$ & $0.95$ & $0.92$ & $0.87$ & $0.81$ \\

$a_{P_1}^{5}$ & $0.94$ & $0.96$ & $0.98$ & $1.$ & $1.$ & $0.99$ & $0.98$ & $0.95$ & $0.92$ & $0.87$ \\

$a_{P_1}^{6}$ & $0.89$ & $0.93$ & $0.96$ & $0.98$ & $0.99$ & $1.$ & $0.99$ & $0.98$ & $0.95$ & $0.91$ \\

$a_{P_1}^{7}$ & $0.84$ & $0.89$ & $0.92$ & $0.95$ & $0.98$ & $0.99$ & $1.$ & $0.99$ & $0.98$ & $0.95$ \\

$a_{P_1}^{8}$ & $0.78$ & $0.83$ & $0.88$ & $0.92$ & $0.95$ & $0.98$ & $0.99$ & $1.$ & $0.99$ & $0.98$ \\

$a_{P_1}^{9}$ & $0.71$ & $0.77$ & $0.82$ & $0.87$ & $0.92$ & $0.95$ & $0.98$ & $0.99$ & $1.$ & $0.99$ \\

$a_{P_1}^{10}$ & $0.63$ & $0.7$ & $0.76$ & $0.81$ & $0.87$ & $0.91$ & $0.95$ & $0.98$ & $0.99$ & $1.$ \\

\end{tabular}}
\caption{\label{tab:corr44} Correlations among the $a_{P_1}$ and $a_{P_1}$ parameters.
}
\end{table}
\FloatBarrier

\begin{table}[h!]
\centering
{\scriptsize
\begin{tabular}{c|cccccccccc}
& $b_f^{1}$ & $b_f^{2}$ & $b_f^{3}$ & $b_f^{4}$ & $b_f^{5}$ & $b_f^{6}$ & $b_f^{7}$ & $b_f^{8}$ & $b_f^{9}$ & $b_f^{10}$ \\
\hline

$a_{P_1}^{1}$ & $-0.6$ & $-0.61$ & $-0.55$ & $-0.48$ & $-0.42$ & $-0.37$ & $-0.33$ & $-0.3$ & $-0.27$ & $-0.25$ \\

$a_{P_1}^{2}$ & $-0.59$ & $-0.61$ & $-0.56$ & $-0.5$ & $-0.43$ & $-0.39$ & $-0.35$ & $-0.32$ & $-0.29$ & $-0.27$ \\

$a_{P_1}^{3}$ & $-0.58$ & $-0.6$ & $-0.56$ & $-0.5$ & $-0.45$ & $-0.4$ & $-0.36$ & $-0.33$ & $-0.31$ & $-0.29$ \\

$a_{P_1}^{4}$ & $-0.56$ & $-0.59$ & $-0.56$ & $-0.51$ & $-0.45$ & $-0.41$ & $-0.38$ & $-0.35$ & $-0.32$ & $-0.31$ \\

$a_{P_1}^{5}$ & $-0.53$ & $-0.58$ & $-0.55$ & $-0.51$ & $-0.46$ & $-0.42$ & $-0.39$ & $-0.36$ & $-0.34$ & $-0.32$ \\

$a_{P_1}^{6}$ & $-0.5$ & $-0.55$ & $-0.54$ & $-0.5$ & $-0.46$ & $-0.42$ & $-0.39$ & $-0.37$ & $-0.35$ & $-0.33$ \\

$a_{P_1}^{7}$ & $-0.47$ & $-0.52$ & $-0.52$ & $-0.49$ & $-0.46$ & $-0.42$ & $-0.4$ & $-0.38$ & $-0.36$ & $-0.34$ \\

$a_{P_1}^{8}$ & $-0.42$ & $-0.49$ & $-0.5$ & $-0.48$ & $-0.45$ & $-0.42$ & $-0.4$ & $-0.38$ & $-0.36$ & $-0.35$ \\

$a_{P_1}^{9}$ & $-0.37$ & $-0.45$ & $-0.47$ & $-0.45$ & $-0.43$ & $-0.41$ & $-0.39$ & $-0.38$ & $-0.36$ & $-0.35$ \\

$a_{P_1}^{10}$ & $-0.32$ & $-0.4$ & $-0.43$ & $-0.43$ & $-0.41$ & $-0.4$ & $-0.38$ & $-0.37$ & $-0.36$ & $-0.35$ \\

\end{tabular}}
\caption{\label{tab:corr45} Correlations among the $a_{P_1}$ and $b_f$ parameters.
}
\end{table}
\FloatBarrier

\begin{table}[h!]
\centering
{\scriptsize
\begin{tabular}{c|cccccccccc}
& $b_g^{1}$ & $b_g^{2}$ & $b_g^{3}$ & $b_g^{4}$ & $b_g^{5}$ & $b_g^{6}$ & $b_g^{7}$ & $b_g^{8}$ & $b_g^{9}$ & $b_g^{10}$  \\
\hline

$a_{P_1}^{1}$ & $-0.42$ & $-0.42$ & $-0.41$ & $-0.41$ & $-0.39$ & $-0.38$ & $-0.36$ & $-0.34$ & $-0.32$ & $-0.3$ \\

$a_{P_1}^{2}$ & $-0.42$ & $-0.42$ & $-0.41$ & $-0.4$ & $-0.39$ & $-0.38$ & $-0.36$ & $-0.34$ & $-0.32$ & $-0.3$ \\

$a_{P_1}^{3}$ & $-0.41$ & $-0.41$ & $-0.41$ & $-0.4$ & $-0.39$ & $-0.37$ & $-0.36$ & $-0.34$ & $-0.32$ & $-0.3$ \\

$a_{P_1}^{4}$ & $-0.41$ & $-0.41$ & $-0.4$ & $-0.39$ & $-0.38$ & $-0.37$ & $-0.35$ & $-0.33$ & $-0.32$ & $-0.3$ \\

$a_{P_1}^{5}$ & $-0.39$ & $-0.39$ & $-0.39$ & $-0.38$ & $-0.37$ & $-0.36$ & $-0.34$ & $-0.32$ & $-0.31$ & $-0.29$ \\

$a_{P_1}^{6}$ & $-0.38$ & $-0.38$ & $-0.37$ & $-0.37$ & $-0.36$ & $-0.34$ & $-0.33$ & $-0.31$ & $-0.3$ & $-0.28$ \\

$a_{P_1}^{7}$ & $-0.35$ & $-0.36$ & $-0.35$ & $-0.35$ & $-0.34$ & $-0.32$ & $-0.31$ & $-0.3$ & $-0.28$ & $-0.27$ \\

$a_{P_1}^{8}$ & $-0.33$ & $-0.33$ & $-0.33$ & $-0.32$ & $-0.31$ & $-0.3$ & $-0.29$ & $-0.28$ & $-0.26$ & $-0.25$ \\

$a_{P_1}^{9}$ & $-0.3$ & $-0.3$ & $-0.3$ & $-0.29$ & $-0.29$ & $-0.28$ & $-0.27$ & $-0.25$ & $-0.24$ & $-0.23$ \\

$a_{P_1}^{10}$ & $-0.27$ & $-0.27$ & $-0.27$ & $-0.26$ & $-0.26$ & $-0.25$ & $-0.24$ & $-0.23$ & $-0.22$ & $-0.21$ \\

\end{tabular}}
\caption{\label{tab:corr46} Correlations among the $a_{P_1}$ and $b_g$ parameters.
}
\end{table}
\FloatBarrier

\begin{table}[h!]
\centering
{\scriptsize
\begin{tabular}{c|cccccccccc}
& $b_{\mathcal{F}_1}^{1}$ & $b_{\mathcal{F}_1}^{2}$ & $b_{\mathcal{F}_1}^{3}$ & $b_{\mathcal{F}_1}^{4}$ & $b_{\mathcal{F}_1}^{5}$ & $b_{\mathcal{F}_1}^{6}$ & $b_{\mathcal{F}_1}^{7}$ & $b_{\mathcal{F}_1}^{8}$ & $b_{\mathcal{F}_1}^{9}$ & $b_{\mathcal{F}_1}^{10}$  \\
\hline

$a_{P_1}^{1}$ & $-0.5$ & $-0.62$ & $-0.69$ & $-0.72$ & $-0.74$ & $-0.74$ & $-0.74$ & $-0.73$ & $-0.71$ & $-0.68$ \\

$a_{P_1}^{2}$ & $-0.48$ & $-0.62$ & $-0.69$ & $-0.73$ & $-0.75$ & $-0.76$ & $-0.76$ & $-0.75$ & $-0.73$ & $-0.7$ \\

$a_{P_1}^{3}$ & $-0.46$ & $-0.6$ & $-0.68$ & $-0.73$ & $-0.75$ & $-0.76$ & $-0.77$ & $-0.76$ & $-0.74$ & $-0.72$ \\

$a_{P_1}^{4}$ & $-0.43$ & $-0.58$ & $-0.67$ & $-0.72$ & $-0.75$ & $-0.76$ & $-0.77$ & $-0.77$ & $-0.75$ & $-0.73$ \\

$a_{P_1}^{5}$ & $-0.4$ & $-0.55$ & $-0.65$ & $-0.7$ & $-0.74$ & $-0.76$ & $-0.77$ & $-0.77$ & $-0.76$ & $-0.74$ \\

$a_{P_1}^{6}$ & $-0.36$ & $-0.52$ & $-0.62$ & $-0.68$ & $-0.72$ & $-0.74$ & $-0.75$ & $-0.76$ & $-0.75$ & $-0.73$ \\

$a_{P_1}^{7}$ & $-0.32$ & $-0.47$ & $-0.58$ & $-0.64$ & $-0.68$ & $-0.71$ & $-0.73$ & $-0.74$ & $-0.73$ & $-0.72$ \\

$a_{P_1}^{8}$ & $-0.27$ & $-0.42$ & $-0.53$ & $-0.6$ & $-0.64$ & $-0.68$ & $-0.7$ & $-0.71$ & $-0.71$ & $-0.7$ \\

$a_{P_1}^{9}$ & $-0.21$ & $-0.36$ & $-0.47$ & $-0.54$ & $-0.6$ & $-0.63$ & $-0.65$ & $-0.67$ & $-0.67$ & $-0.66$ \\

$a_{P_1}^{10}$ & $-0.16$ & $-0.3$ & $-0.41$ & $-0.48$ & $-0.54$ & $-0.58$ & $-0.6$ & $-0.62$ & $-0.62$ & $-0.62$ \\

\end{tabular}}
\caption{\label{tab:corr47} Correlations among the $a_{P_1}$ and $b_{\mathcal{F}_1}$ parameters.
}
\end{table}
\FloatBarrier

\begin{table}[h!]
\centering
{\scriptsize
\begin{tabular}{c|cccccccccc}
& $b_{P_1}^{1}$ & $b_{P_1}^{2}$ & $b_{P_1}^{3}$ & $b_{P_1}^{4}$ & $b_{P_1}^{5}$ & $b_{P_1}^{6}$ & $b_{P_1}^{7}$ & $b_{P_1}^{8}$ & $b_{P_1}^{9}$ & $b_{P_1}^{10}$  \\
\hline

$a_{P_1}^{1}$ & $-0.99$ & $-0.98$ & $-0.97$ & $-0.95$ & $-0.93$ & $-0.9$ & $-0.87$ & $-0.83$ & $-0.79$ & $-0.74$ \\

$a_{P_1}^{2}$ & $-0.98$ & $-0.98$ & $-0.98$ & $-0.97$ & $-0.95$ & $-0.93$ & $-0.9$ & $-0.87$ & $-0.83$ & $-0.79$ \\

$a_{P_1}^{3}$ & $-0.96$ & $-0.97$ & $-0.98$ & $-0.97$ & $-0.96$ & $-0.95$ & $-0.92$ & $-0.9$ & $-0.87$ & $-0.84$ \\

$a_{P_1}^{4}$ & $-0.94$ & $-0.96$ & $-0.97$ & $-0.97$ & $-0.97$ & $-0.96$ & $-0.95$ & $-0.93$ & $-0.9$ & $-0.87$ \\

$a_{P_1}^{5}$ & $-0.91$ & $-0.93$ & $-0.95$ & $-0.96$ & $-0.97$ & $-0.96$ & $-0.96$ & $-0.94$ & $-0.93$ & $-0.91$ \\

$a_{P_1}^{6}$ & $-0.86$ & $-0.89$ & $-0.92$ & $-0.94$ & $-0.95$ & $-0.96$ & $-0.96$ & $-0.95$ & $-0.94$ & $-0.93$ \\

$a_{P_1}^{7}$ & $-0.81$ & $-0.85$ & $-0.88$ & $-0.91$ & $-0.93$ & $-0.94$ & $-0.95$ & $-0.95$ & $-0.95$ & $-0.94$ \\

$a_{P_1}^{8}$ & $-0.74$ & $-0.79$ & $-0.83$ & $-0.86$ & $-0.89$ & $-0.92$ & $-0.93$ & $-0.94$ & $-0.94$ & $-0.94$ \\

$a_{P_1}^{9}$ & $-0.67$ & $-0.72$ & $-0.77$ & $-0.81$ & $-0.85$ & $-0.88$ & $-0.9$ & $-0.92$ & $-0.93$ & $-0.93$ \\

$a_{P_1}^{10}$ & $-0.58$ & $-0.64$ & $-0.7$ & $-0.75$ & $-0.79$ & $-0.83$ & $-0.86$ & $-0.88$ & $-0.9$ & $-0.91$ \\

\end{tabular}}
\caption{\label{tab:corr48} Correlations among the $a_{P_1}$ and $b_{P_1}$ parameters.
}
\end{table}
\FloatBarrier

\begin{table}[h!]
\centering
{\footnotesize
\begin{tabular}{c|cccccccccc}
& $b_f^{1}$ & $b_f^{2}$ & $b_f^{3}$ & $b_f^{4}$ & $b_f^{5}$ & $b_f^{6}$ & $b_f^{7}$ & $b_f^{8}$ & $b_f^{9}$ & $b_f^{10}$ \\
\hline

$b_f^{1}$ & $1.$ & $0.94$ & $0.79$ & $0.64$ & $0.52$ & $0.42$ & $0.35$ & $0.29$ & $0.25$ & $0.21$ \\

$b_f^{2}$ & $0.94$ & $1.$ & $0.96$ & $0.87$ & $0.78$ & $0.71$ & $0.65$ & $0.6$ & $0.57$ & $0.53$ \\

$b_f^{3}$ & $0.79$ & $0.96$ & $1.$ & $0.98$ & $0.93$ & $0.89$ & $0.85$ & $0.81$ & $0.78$ & $0.76$ \\

$b_f^{4}$ & $0.64$ & $0.87$ & $0.98$ & $1.$ & $0.99$ & $0.97$ & $0.94$ & $0.92$ & $0.9$ & $0.88$ \\

$b_f^{5}$ & $0.52$ & $0.78$ & $0.93$ & $0.99$ & $1.$ & $0.99$ & $0.98$ & $0.97$ & $0.96$ & $0.95$ \\

$b_f^{6}$ & $0.42$ & $0.71$ & $0.89$ & $0.97$ & $0.99$ & $1.$ & $1.$ & $0.99$ & $0.98$ & $0.98$ \\

$b_f^{7}$ & $0.35$ & $0.65$ & $0.85$ & $0.94$ & $0.98$ & $1.$ & $1.$ & $1.$ & $0.99$ & $0.99$ \\

$b_f^{8}$ & $0.29$ & $0.6$ & $0.81$ & $0.92$ & $0.97$ & $0.99$ & $1.$ & $1.$ & $1.$ & $1.$ \\

$b_f^{9}$ & $0.25$ & $0.57$ & $0.78$ & $0.9$ & $0.96$ & $0.98$ & $0.99$ & $1.$ & $1.$ & $1.$ \\

$b_f^{10}$ & $0.21$ & $0.53$ & $0.76$ & $0.88$ & $0.95$ & $0.98$ & $0.99$ & $1.$ & $1.$ & $1.$ \\

\end{tabular}}
\caption{\label{tab:corr55} Correlations among the $b_f$ and $b_f$ parameters.
}
\end{table}
\FloatBarrier

\begin{table}[h!]
\centering
{\footnotesize
\begin{tabular}{c|cccccccccc}
& $b_g^{1}$ & $b_g^{2}$ & $b_g^{3}$ & $b_g^{4}$ & $b_g^{5}$ & $b_g^{6}$ & $b_g^{7}$ & $b_g^{8}$ & $b_g^{9}$ & $b_g^{10}$  \\
\hline

$b_f^{1}$ & $0.29$ & $0.29$ & $0.28$ & $0.27$ & $0.26$ & $0.25$ & $0.23$ & $0.22$ & $0.2$ & $0.19$ \\

$b_f^{2}$ & $0.36$ & $0.36$ & $0.35$ & $0.34$ & $0.33$ & $0.31$ & $0.29$ & $0.28$ & $0.26$ & $0.24$ \\

$b_f^{3}$ & $0.38$ & $0.38$ & $0.37$ & $0.36$ & $0.35$ & $0.33$ & $0.32$ & $0.3$ & $0.28$ & $0.26$ \\

$b_f^{4}$ & $0.38$ & $0.38$ & $0.37$ & $0.36$ & $0.35$ & $0.33$ & $0.32$ & $0.3$ & $0.28$ & $0.26$ \\

$b_f^{5}$ & $0.37$ & $0.36$ & $0.36$ & $0.35$ & $0.34$ & $0.32$ & $0.31$ & $0.29$ & $0.27$ & $0.25$ \\

$b_f^{6}$ & $0.35$ & $0.35$ & $0.34$ & $0.34$ & $0.32$ & $0.31$ & $0.3$ & $0.28$ & $0.26$ & $0.25$ \\

$b_f^{7}$ & $0.34$ & $0.34$ & $0.33$ & $0.32$ & $0.31$ & $0.3$ & $0.28$ & $0.27$ & $0.25$ & $0.24$ \\

$b_f^{8}$ & $0.33$ & $0.32$ & $0.32$ & $0.31$ & $0.3$ & $0.29$ & $0.28$ & $0.26$ & $0.25$ & $0.23$ \\

$b_f^{9}$ & $0.32$ & $0.32$ & $0.31$ & $0.3$ & $0.29$ & $0.28$ & $0.27$ & $0.25$ & $0.24$ & $0.22$ \\

$b_f^{10}$ & $0.31$ & $0.31$ & $0.3$ & $0.3$ & $0.29$ & $0.27$ & $0.26$ & $0.25$ & $0.23$ & $0.22$ \\

\end{tabular}}
\caption{\label{tab:corr56} Correlations among the $b_f$ and $b_g$ parameters.
}
\end{table}
\FloatBarrier

\begin{table}[h!]
\centering
{\footnotesize
\begin{tabular}{c|cccccccccc}
& $b_{\mathcal{F}_1}^{1}$ & $b_{\mathcal{F}_1}^{2}$ & $b_{\mathcal{F}_1}^{3}$ & $b_{\mathcal{F}_1}^{4}$ & $b_{\mathcal{F}_1}^{5}$ & $b_{\mathcal{F}_1}^{6}$ & $b_{\mathcal{F}_1}^{7}$ & $b_{\mathcal{F}_1}^{8}$ & $b_{\mathcal{F}_1}^{9}$ & $b_{\mathcal{F}_1}^{10}$  \\
\hline

$b_f^{1}$ & $0.7$ & $0.69$ & $0.64$ & $0.59$ & $0.54$ & $0.5$ & $0.46$ & $0.43$ & $0.4$ & $0.37$ \\

$b_f^{2}$ & $0.63$ & $0.65$ & $0.63$ & $0.59$ & $0.55$ & $0.51$ & $0.48$ & $0.45$ & $0.41$ & $0.38$ \\

$b_f^{3}$ & $0.51$ & $0.55$ & $0.55$ & $0.53$ & $0.5$ & $0.47$ & $0.44$ & $0.41$ & $0.38$ & $0.35$ \\

$b_f^{4}$ & $0.39$ & $0.45$ & $0.46$ & $0.46$ & $0.44$ & $0.42$ & $0.39$ & $0.37$ & $0.34$ & $0.31$ \\

$b_f^{5}$ & $0.3$ & $0.36$ & $0.39$ & $0.39$ & $0.38$ & $0.37$ & $0.35$ & $0.33$ & $0.3$ & $0.27$ \\

$b_f^{6}$ & $0.23$ & $0.3$ & $0.33$ & $0.34$ & $0.34$ & $0.33$ & $0.31$ & $0.29$ & $0.27$ & $0.24$ \\

$b_f^{7}$ & $0.17$ & $0.25$ & $0.28$ & $0.3$ & $0.3$ & $0.29$ & $0.28$ & $0.26$ & $0.24$ & $0.22$ \\

$b_f^{8}$ & $0.13$ & $0.21$ & $0.25$ & $0.27$ & $0.27$ & $0.27$ & $0.26$ & $0.24$ & $0.22$ & $0.2$ \\

$b_f^{9}$ & $0.1$ & $0.18$ & $0.22$ & $0.24$ & $0.25$ & $0.25$ & $0.24$ & $0.22$ & $0.2$ & $0.18$ \\

$b_f^{10}$ & $0.07$ & $0.15$ & $0.2$ & $0.22$ & $0.23$ & $0.23$ & $0.22$ & $0.21$ & $0.19$ & $0.17$ \\

\end{tabular}}
\caption{\label{tab:corr57} Correlations among the $b_f$ and $b_{\mathcal{F}_1}$ parameters.
}
\end{table}
\FloatBarrier

\begin{table}[h!]
\centering
{\scriptsize
\begin{tabular}{c|cccccccccc}
& $b_{P_1}^{1}$ & $b_{P_1}^{2}$ & $b_{P_1}^{3}$ & $b_{P_1}^{4}$ & $b_{P_1}^{5}$ & $b_{P_1}^{6}$ & $b_{P_1}^{7}$ & $b_{P_1}^{8}$ & $b_{P_1}^{9}$ & $b_{P_1}^{10}$  \\
\hline

$b_f^{1}$ & $0.61$ & $0.61$ & $0.6$ & $0.59$ & $0.57$ & $0.55$ & $0.52$ & $0.5$ & $0.47$ & $0.44$ \\

$b_f^{2}$ & $0.61$ & $0.61$ & $0.61$ & $0.6$ & $0.59$ & $0.57$ & $0.55$ & $0.53$ & $0.51$ & $0.49$ \\

$b_f^{3}$ & $0.54$ & $0.55$ & $0.55$ & $0.55$ & $0.54$ & $0.54$ & $0.52$ & $0.51$ & $0.49$ & $0.48$ \\

$b_f^{4}$ & $0.46$ & $0.47$ & $0.48$ & $0.48$ & $0.48$ & $0.48$ & $0.47$ & $0.46$ & $0.45$ & $0.44$ \\

$b_f^{5}$ & $0.4$ & $0.41$ & $0.42$ & $0.42$ & $0.43$ & $0.43$ & $0.42$ & $0.42$ & $0.41$ & $0.4$ \\

$b_f^{6}$ & $0.34$ & $0.35$ & $0.37$ & $0.37$ & $0.38$ & $0.38$ & $0.38$ & $0.38$ & $0.38$ & $0.37$ \\

$b_f^{7}$ & $0.3$ & $0.31$ & $0.33$ & $0.33$ & $0.34$ & $0.35$ & $0.35$ & $0.35$ & $0.35$ & $0.35$ \\

$b_f^{8}$ & $0.27$ & $0.28$ & $0.29$ & $0.3$ & $0.31$ & $0.32$ & $0.32$ & $0.33$ & $0.33$ & $0.32$ \\

$b_f^{9}$ & $0.24$ & $0.25$ & $0.27$ & $0.28$ & $0.29$ & $0.3$ & $0.3$ & $0.31$ & $0.31$ & $0.31$ \\

$b_f^{10}$ & $0.22$ & $0.23$ & $0.25$ & $0.26$ & $0.27$ & $0.28$ & $0.28$ & $0.29$ & $0.29$ & $0.29$ \\

\end{tabular}}
\caption{\label{tab:corr58} Correlations among the $b_f$ and $b_{P_1}$ parameters.
}
\end{table}
\FloatBarrier

\begin{table}[h!]
\centering
{\footnotesize
\begin{tabular}{c|cccccccccc}
& $b_g^{1}$ & $b_g^{2}$ & $b_g^{3}$ & $b_g^{4}$ & $b_g^{5}$ & $b_g^{6}$ & $b_g^{7}$ & $b_g^{8}$ & $b_g^{9}$ & $b_g^{10}$  \\
\hline

$b_g^{1}$ & $1.$ & $0.99$ & $0.98$ & $0.96$ & $0.92$ & $0.88$ & $0.84$ & $0.79$ & $0.75$ & $0.7$ \\

$b_g^{2}$ & $0.99$ & $1.$ & $0.99$ & $0.98$ & $0.96$ & $0.93$ & $0.89$ & $0.85$ & $0.81$ & $0.77$ \\

$b_g^{3}$ & $0.98$ & $0.99$ & $1.$ & $1.$ & $0.98$ & $0.96$ & $0.93$ & $0.9$ & $0.87$ & $0.83$ \\

$b_g^{4}$ & $0.96$ & $0.98$ & $1.$ & $1.$ & $1.$ & $0.98$ & $0.96$ & $0.94$ & $0.91$ & $0.88$ \\

$b_g^{5}$ & $0.92$ & $0.96$ & $0.98$ & $1.$ & $1.$ & $1.$ & $0.98$ & $0.97$ & $0.95$ & $0.92$ \\

$b_g^{6}$ & $0.88$ & $0.93$ & $0.96$ & $0.98$ & $1.$ & $1.$ & $1.$ & $0.99$ & $0.97$ & $0.95$ \\

$b_g^{7}$ & $0.84$ & $0.89$ & $0.93$ & $0.96$ & $0.98$ & $1.$ & $1.$ & $1.$ & $0.99$ & $0.98$ \\

$b_g^{8}$ & $0.79$ & $0.85$ & $0.9$ & $0.94$ & $0.97$ & $0.99$ & $1.$ & $1.$ & $1.$ & $0.99$ \\

$b_g^{9}$ & $0.75$ & $0.81$ & $0.87$ & $0.91$ & $0.95$ & $0.97$ & $0.99$ & $1.$ & $1.$ & $1.$ \\

$b_g^{10}$ & $0.7$ & $0.77$ & $0.83$ & $0.88$ & $0.92$ & $0.95$ & $0.98$ & $0.99$ & $1.$ & $1.$ \\

\end{tabular}}
\caption{\label{tab:corr66} Correlations among the $b_g$ and $b_g$ parameters.
}
\end{table}
\FloatBarrier

\begin{table}[h!]
\centering
{\footnotesize
\begin{tabular}{c|cccccccccc}
& $b_{\mathcal{F}_1}^{1}$ & $b_{\mathcal{F}_1}^{2}$ & $b_{\mathcal{F}_1}^{3}$ & $b_{\mathcal{F}_1}^{4}$ & $b_{\mathcal{F}_1}^{5}$ & $b_{\mathcal{F}_1}^{6}$ & $b_{\mathcal{F}_1}^{7}$ & $b_{\mathcal{F}_1}^{8}$ & $b_{\mathcal{F}_1}^{9}$ & $b_{\mathcal{F}_1}^{10}$  \\
\hline

$b_g^{1}$ & $0.14$ & $0.21$ & $0.26$ & $0.29$ & $0.31$ & $0.32$ & $0.33$ & $0.33$ & $0.33$ & $0.32$ \\

$b_g^{2}$ & $0.14$ & $0.21$ & $0.26$ & $0.29$ & $0.31$ & $0.33$ & $0.33$ & $0.33$ & $0.33$ & $0.32$ \\

$b_g^{3}$ & $0.13$ & $0.21$ & $0.26$ & $0.29$ & $0.31$ & $0.32$ & $0.33$ & $0.33$ & $0.33$ & $0.32$ \\

$b_g^{4}$ & $0.13$ & $0.2$ & $0.26$ & $0.29$ & $0.31$ & $0.32$ & $0.33$ & $0.33$ & $0.32$ & $0.32$ \\

$b_g^{5}$ & $0.12$ & $0.2$ & $0.25$ & $0.28$ & $0.3$ & $0.31$ & $0.32$ & $0.32$ & $0.32$ & $0.31$ \\

$b_g^{6}$ & $0.12$ & $0.19$ & $0.24$ & $0.27$ & $0.29$ & $0.3$ & $0.31$ & $0.31$ & $0.31$ & $0.3$ \\

$b_g^{7}$ & $0.11$ & $0.18$ & $0.23$ & $0.26$ & $0.28$ & $0.29$ & $0.3$ & $0.3$ & $0.3$ & $0.29$ \\

$b_g^{8}$ & $0.11$ & $0.17$ & $0.22$ & $0.25$ & $0.27$ & $0.28$ & $0.29$ & $0.29$ & $0.28$ & $0.28$ \\

$b_g^{9}$ & $0.1$ & $0.17$ & $0.21$ & $0.24$ & $0.26$ & $0.27$ & $0.27$ & $0.27$ & $0.27$ & $0.26$ \\

$b_g^{10}$ & $0.09$ & $0.16$ & $0.2$ & $0.23$ & $0.24$ & $0.25$ & $0.26$ & $0.26$ & $0.26$ & $0.25$ \\

\end{tabular}}
\caption{\label{tab:corr67} Correlations among the $b_g$ and $b_{\mathcal{F}_1}$ parameters.
}
\end{table}
\FloatBarrier

\begin{table}[h!]
\centering
{\scriptsize
\begin{tabular}{c|cccccccccc}
& $b_{P_1}^{1}$ & $b_{P_1}^{2}$ & $b_{P_1}^{3}$ & $b_{P_1}^{4}$ & $b_{P_1}^{5}$ & $b_{P_1}^{6}$ & $b_{P_1}^{7}$ & $b_{P_1}^{8}$ & $b_{P_1}^{9}$ & $b_{P_1}^{10}$  \\
\hline

$b_g^{1}$ & $0.42$ & $0.42$ & $0.41$ & $0.4$ & $0.39$ & $0.38$ & $0.37$ & $0.35$ & $0.33$ & $0.32$ \\

$b_g^{2}$ & $0.42$ & $0.42$ & $0.41$ & $0.41$ & $0.4$ & $0.38$ & $0.37$ & $0.35$ & $0.34$ & $0.32$ \\

$b_g^{3}$ & $0.41$ & $0.41$ & $0.41$ & $0.4$ & $0.39$ & $0.38$ & $0.37$ & $0.35$ & $0.34$ & $0.32$ \\

$b_g^{4}$ & $0.4$ & $0.4$ & $0.4$ & $0.39$ & $0.39$ & $0.37$ & $0.36$ & $0.35$ & $0.33$ & $0.31$ \\

$b_g^{5}$ & $0.39$ & $0.39$ & $0.39$ & $0.38$ & $0.38$ & $0.36$ & $0.35$ & $0.34$ & $0.32$ & $0.31$ \\

$b_g^{6}$ & $0.38$ & $0.38$ & $0.37$ & $0.37$ & $0.36$ & $0.35$ & $0.34$ & $0.33$ & $0.31$ & $0.3$ \\

$b_g^{7}$ & $0.36$ & $0.36$ & $0.36$ & $0.35$ & $0.35$ & $0.34$ & $0.33$ & $0.32$ & $0.3$ & $0.29$ \\

$b_g^{8}$ & $0.34$ & $0.34$ & $0.34$ & $0.34$ & $0.33$ & $0.32$ & $0.31$ & $0.3$ & $0.29$ & $0.28$ \\

$b_g^{9}$ & $0.32$ & $0.32$ & $0.32$ & $0.32$ & $0.31$ & $0.31$ & $0.3$ & $0.29$ & $0.28$ & $0.26$ \\

$b_g^{10}$ & $0.3$ & $0.31$ & $0.31$ & $0.3$ & $0.3$ & $0.29$ & $0.28$ & $0.27$ & $0.26$ & $0.25$ \\

\end{tabular}}
\caption{\label{tab:corr68} Correlations among the $b_g$ and $b_{P_1}$ parameters.
}
\end{table}
\FloatBarrier

\begin{table}[h!]
\centering
{\footnotesize
\begin{tabular}{c|cccccccccc}
& $b_{\mathcal{F}_1}^{1}$ & $b_{\mathcal{F}_1}^{2}$ & $b_{\mathcal{F}_1}^{3}$ & $b_{\mathcal{F}_1}^{4}$ & $b_{\mathcal{F}_1}^{5}$ & $b_{\mathcal{F}_1}^{6}$ & $b_{\mathcal{F}_1}^{7}$ & $b_{\mathcal{F}_1}^{8}$ & $b_{\mathcal{F}_1}^{9}$ & $b_{\mathcal{F}_1}^{10}$  \\
\hline

$b_{\mathcal{F}_1}^{1}$ & $1.$ & $0.95$ & $0.86$ & $0.76$ & $0.68$ & $0.61$ & $0.55$ & $0.5$ & $0.46$ & $0.42$ \\

$b_{\mathcal{F}_1}^{2}$ & $0.95$ & $1.$ & $0.97$ & $0.92$ & $0.86$ & $0.8$ & $0.75$ & $0.69$ & $0.64$ & $0.58$ \\

$b_{\mathcal{F}_1}^{3}$ & $0.86$ & $0.97$ & $1.$ & $0.98$ & $0.95$ & $0.91$ & $0.87$ & $0.82$ & $0.76$ & $0.7$ \\

$b_{\mathcal{F}_1}^{4}$ & $0.76$ & $0.92$ & $0.98$ & $1.$ & $0.99$ & $0.97$ & $0.93$ & $0.89$ & $0.84$ & $0.79$ \\

$b_{\mathcal{F}_1}^{5}$ & $0.68$ & $0.86$ & $0.95$ & $0.99$ & $1.$ & $0.99$ & $0.97$ & $0.94$ & $0.9$ & $0.85$ \\

$b_{\mathcal{F}_1}^{6}$ & $0.61$ & $0.8$ & $0.91$ & $0.97$ & $0.99$ & $1.$ & $0.99$ & $0.97$ & $0.95$ & $0.9$ \\

$b_{\mathcal{F}_1}^{7}$ & $0.55$ & $0.75$ & $0.87$ & $0.93$ & $0.97$ & $0.99$ & $1.$ & $0.99$ & $0.98$ & $0.95$ \\

$b_{\mathcal{F}_1}^{8}$ & $0.5$ & $0.69$ & $0.82$ & $0.89$ & $0.94$ & $0.97$ & $0.99$ & $1.$ & $0.99$ & $0.98$ \\

$b_{\mathcal{F}_1}^{9}$ & $0.46$ & $0.64$ & $0.76$ & $0.84$ & $0.9$ & $0.95$ & $0.98$ & $0.99$ & $1.$ & $0.99$ \\

$b_{\mathcal{F}_1}^{10}$ & $0.42$ & $0.58$ & $0.7$ & $0.79$ & $0.85$ & $0.9$ & $0.95$ & $0.98$ & $0.99$ & $1.$ \\

\end{tabular}}
\caption{\label{tab:corr77} Correlations among the $b_{\mathcal{F}_1}$ and $b_{\mathcal{F}_1}$ parameters.
}
\end{table}
\FloatBarrier

\begin{table}[h!]
\centering
{\scriptsize
\begin{tabular}{c|cccccccccc}
& $b_{P_1}^{1}$ & $b_{P_1}^{2}$ & $b_{P_1}^{3}$ & $b_{P_1}^{4}$ & $b_{P_1}^{5}$ & $b_{P_1}^{6}$ & $b_{P_1}^{7}$ & $b_{P_1}^{8}$ & $b_{P_1}^{9}$ & $b_{P_1}^{10}$  \\
\hline

$b_{\mathcal{F}_1}^{1}$ & $0.57$ & $0.56$ & $0.55$ & $0.54$ & $0.52$ & $0.5$ & $0.48$ & $0.46$ & $0.43$ & $0.4$ \\

$b_{\mathcal{F}_1}^{2}$ & $0.71$ & $0.71$ & $0.71$ & $0.7$ & $0.69$ & $0.67$ & $0.65$ & $0.63$ & $0.61$ & $0.58$ \\

$b_{\mathcal{F}_1}^{3}$ & $0.78$ & $0.79$ & $0.79$ & $0.79$ & $0.79$ & $0.78$ & $0.76$ & $0.74$ & $0.72$ & $0.7$ \\

$b_{\mathcal{F}_1}^{4}$ & $0.81$ & $0.83$ & $0.84$ & $0.84$ & $0.84$ & $0.84$ & $0.83$ & $0.81$ & $0.79$ & $0.77$ \\

$b_{\mathcal{F}_1}^{5}$ & $0.82$ & $0.84$ & $0.86$ & $0.87$ & $0.87$ & $0.87$ & $0.87$ & $0.85$ & $0.84$ & $0.82$ \\

$b_{\mathcal{F}_1}^{6}$ & $0.82$ & $0.85$ & $0.87$ & $0.88$ & $0.89$ & $0.89$ & $0.89$ & $0.88$ & $0.86$ & $0.85$ \\

$b_{\mathcal{F}_1}^{7}$ & $0.82$ & $0.85$ & $0.87$ & $0.89$ & $0.9$ & $0.9$ & $0.9$ & $0.89$ & $0.88$ & $0.86$ \\

$b_{\mathcal{F}_1}^{8}$ & $0.8$ & $0.83$ & $0.86$ & $0.88$ & $0.89$ & $0.9$ & $0.9$ & $0.89$ & $0.88$ & $0.87$ \\

$b_{\mathcal{F}_1}^{9}$ & $0.78$ & $0.81$ & $0.84$ & $0.86$ & $0.87$ & $0.88$ & $0.88$ & $0.88$ & $0.87$ & $0.86$ \\

$b_{\mathcal{F}_1}^{10}$ & $0.75$ & $0.78$ & $0.81$ & $0.83$ & $0.85$ & $0.86$ & $0.86$ & $0.86$ & $0.85$ & $0.84$ \\

\end{tabular}}
\caption{\label{tab:corr78} Correlations among the $b_{\mathcal{F}_1}$ and $b_{P_1}$ parameters.
}
\end{table}
\FloatBarrier

\begin{table}[h!]
\centering
{\scriptsize
\begin{tabular}{c|cccccccccc}
& $b_{P_1}^{1}$ & $b_{P_1}^{2}$ & $b_{P_1}^{3}$ & $b_{P_1}^{4}$ & $b_{P_1}^{5}$ & $b_{P_1}^{6}$ & $b_{P_1}^{7}$ & $b_{P_1}^{8}$ & $b_{P_1}^{9}$ & $b_{P_1}^{10}$  \\
\hline

$b_{P_1}^{1}$ & $1.$ & $1.$ & $0.99$ & $0.97$ & $0.94$ & $0.91$ & $0.88$ & $0.84$ & $0.8$ & $0.76$ \\

$b_{P_1}^{2}$ & $1.$ & $1.$ & $1.$ & $0.99$ & $0.97$ & $0.94$ & $0.92$ & $0.88$ & $0.85$ & $0.81$ \\

$b_{P_1}^{3}$ & $0.99$ & $1.$ & $1.$ & $1.$ & $0.99$ & $0.97$ & $0.95$ & $0.92$ & $0.89$ & $0.86$ \\

$b_{P_1}^{4}$ & $0.97$ & $0.99$ & $1.$ & $1.$ & $1.$ & $0.99$ & $0.97$ & $0.95$ & $0.93$ & $0.9$ \\

$b_{P_1}^{5}$ & $0.94$ & $0.97$ & $0.99$ & $1.$ & $1.$ & $1.$ & $0.99$ & $0.97$ & $0.95$ & $0.93$ \\

$b_{P_1}^{6}$ & $0.91$ & $0.94$ & $0.97$ & $0.99$ & $1.$ & $1.$ & $1.$ & $0.99$ & $0.97$ & $0.96$ \\

$b_{P_1}^{7}$ & $0.88$ & $0.92$ & $0.95$ & $0.97$ & $0.99$ & $1.$ & $1.$ & $1.$ & $0.99$ & $0.98$ \\

$b_{P_1}^{8}$ & $0.84$ & $0.88$ & $0.92$ & $0.95$ & $0.97$ & $0.99$ & $1.$ & $1.$ & $1.$ & $0.99$ \\

$b_{P_1}^{9}$ & $0.8$ & $0.85$ & $0.89$ & $0.93$ & $0.95$ & $0.97$ & $0.99$ & $1.$ & $1.$ & $1.$ \\

$b_{P_1}^{10}$ & $0.76$ & $0.81$ & $0.86$ & $0.9$ & $0.93$ & $0.96$ & $0.98$ & $0.99$ & $1.$ & $1.$ \\

\end{tabular}}
\caption{\label{tab:corr88} Correlations among the $b_{P_1}$ and $b_{P_1}$ parameters.
}
\end{table}
\FloatBarrier

\bibliography{BIB}

\end{document}